\documentclass[%
 reprint,
nofootinbib,
 amsmath,amssymb,
 aps,
]{revtex4-1}

\usepackage{graphicx}
\usepackage{dcolumn}
\usepackage{bm}
\usepackage{lipsum}
\usepackage[utf8]{inputenc}
\usepackage{verbatim} 
\usepackage{float}

\begin{document}

\title{On the properties of a Static and a Stationary Charged Black Hole in $f(R)$ Gravity.}

\author{Carlos Conde}
 \email{cacondeo@unal.edu.co}
\affiliation{Universidad Nacional de Colombia. Sede Bogotá. Facultad de Ciencias. Observatorio Astronómico Nacional. Ciudad Universitaria. Bogota, Colombia.
}

\author{Cristian Galvis}
 \homepage{cagalvisf@unal.edu.co}
\affiliation{
Universidad Nacional de Colombia. Sede Bogotá. Facultad de Ciencias. Departamento de Física. Ciudad Universitaria. Bogota, Colombia.
}%
\author{Eduard Larra\~{n}aga}
 \homepage{ealarranaga@unal.edu.co}
\affiliation{%
 Universidad Nacional de Colombia. Sede Bogotá. Facultad de Ciencias. Observatorio Astronómico Nacional. Ciudad Universitaria. Bogota, Colombia.}%

\date{\today}

\begin{abstract}
 In a recent paper, Nashed and Capozziello presented a new class of charged, spherically symmetric black hole solutions of $f(R)$ gravity with an asymptotic  flat or (anti-)de Sitter behavior. These metrics depend on a dimensional parameter $\alpha$ and are interesting because they cannot reduce to general relativity solutions. In this paper, we present a corrected study of their physical and thermodynamic properties and generalize these solutions to obtain a new set of stationary, axisymmetric black holes in the $f(R)$ scenario. Some of our results show that the entropy is always positive within the allowed values of parameter $\alpha$ and due to the well behaved quantities such as the Gibbs free energy, we conclude that there is no such a phase transition as discussed in the work of Nashed and Capozziello. We also study the geodesics in these spacetimes and particularly, the stability of the circular orbits to obtain the radius of the Innermost Stable Circular Orbit.

\end{abstract}

\pacs{04.70.Dy, 04.70.Bw, 11.25.-w}

\maketitle

\section{Introduction}

Recently, Nashed and Capozziello \cite{Nashed2019} found a new class of charged spherically symmetric black hole solutions with a flat or (anti-)de Sitter asymptotic behavior in the context of the $f(R)$ gravitational scenario with the particular function $f(R) = R- 2\alpha \sqrt{R-2\Lambda}$, where $\Lambda$ is the cosmological constant. Although the reported metrics are indeed solutions of the Maxwell-$f(R)$ field equations, the physical properties, thermodynamics and stability study presented by the authors in \cite{Nashed2019} have some errors that we will correct in this paper. 

We will also generalize the Nashed-Capozziello metric to obtain a new class of stationary, axisymmetric black hole solution and calculate some of their physical properties.

The work is organized as follows: in Section II we present the charged, spherically symmetric black hole solutions found by Nashed and Capozziello in \cite{Nashed2019} and calculate the correct radius of the event horizon as well as the values of two of the curvature invariants. In section III we introduce the correct thermodynamic quantities such as Hawking temperature, entropy and Gibbs free energy of this black hole to show that there is no indication of phase transitions for these solutions. Section IV is dedicated to study the geodesics and geodesic deviation stability for circular orbits. The obtained equations permit us to find the radius of the Innermost Stable Circular Orbit (ISCO) for these black holes. In Section V we introduce the stationary, axisymmetric solution and present its horizon structure, Hawking temperature and ergosphere. Finally, Section VI presents some conclusions.

\vspace{-0.5cm}

\section{The Static Charged Black Hole Solutions}

The model of gravity used to obtain the black hole solutions is given by the action
\begin{equation}
    \mathcal{S} = \mathcal{S}_g + \mathcal{S}_{EM}
\end{equation}
where $\mathcal{S}_{EM}$ is the electromagnetic field action and
\begin{equation}
    \mathcal{S}_g = \frac{1}{2\kappa}\int \sqrt{-g}\left[ f(R) - \Lambda \right]
\end{equation}
is the gravitational action with $\Lambda$ the cosmological constant, $R$ the Ricci scalar, $g$ the determinant of the metric and $\kappa$ the gravitational constant. The resulting Maxwell-$f(R)$ field equations are
\begin{eqnarray}
    R_{\mu \nu} f'(R) -\frac{1}{2} g_{\mu \nu} f(R) -2 g_{\mu \nu} \Lambda & &\notag \\ 
    + g_{\mu \nu} \Box f'(R) - \nabla_{\mu} \nabla_{\nu} f'(R) &=& 8\pi T_{\mu \nu} \\
    \partial_{\nu} \left( \sqrt{-g} F^{\mu \nu} \right) &=& 0
\end{eqnarray}
where $R_{\mu \nu}$ is the Ricci tensor, $F^{\mu \nu}$ is the electromagnetic field strength tensor, $f'(R) = \frac{df}{dR}$ and
\begin{equation}
    T_{\mu \nu} = \frac{1}{4\pi} \left[ g_{\alpha \beta} F_{\mu}^{\,\, \alpha} F_{\nu}^{\,\, \beta} -\frac{1}{4}g_{\mu \nu} F^{\alpha \beta} F_{\alpha \beta} \right]
\end{equation}
The particular model considered to obtain the solutions is given by the function
\begin{equation}
    f(R) = R - 2\alpha \sqrt{R-8\Lambda},
\end{equation}
where $\alpha$ is a dimensional parameter with positive values.

\subsection{The Asymptotically Flat Black Hole}
The first solution presented in \cite{Nashed2019} represents a charged, spherically symmetric black hole in the absence of cosmological constant and therefore, the space-time is asymptotically flat. The line element is
\begin{equation}
    ds^2 = B(r) dt^2 - \frac{dr^2}{B(r)} - r^2 (d\theta^2 + \sin^2 \theta d\phi^2) \label{eq:LineElement}
\end{equation}
with 
\begin{equation}
    B(r) = \frac{1}{2} - \frac{1}{3\alpha r} + \frac{1}{3\alpha r^2} \label{eq:FlatBH}
\end{equation}
and the gauge potential is given by
\begin{equation}
    A^\mu (r) = \frac{1}{\sqrt{3\alpha} r} dt.
\end{equation}
From these equations it is clear that the parameter $\alpha$ cannot be zero and therefore this solution cannot reduce to a metric in general relativity, e.g. Schwarzschild's or Reissner-Nördstrom's solutions. Ricci's curvature scalar $R = g^{\mu \nu} R_{\mu \nu} $ and Kretchamnn scalar $  K = R_{\alpha \beta \mu \nu} R^{\alpha \beta \mu \nu}$ for this new solution are  

\begin{eqnarray}
    R &=& \frac{1}{r^2} \\
    K &=& \frac{9\alpha^2 r^4 + 12 \alpha r^3 - 12 (\alpha -1) r^2 - 48 r +56}{9 \alpha^2 r^8}.
\end{eqnarray}

These invariants show that the point $r=0$ is an essential singularity and stress out that $\alpha \neq 0$, i.e. the solution cannot reduce to general relativity. \\

The event horizon of a spherically symmetric black hole with a line element as that defined in Eq. (\ref{eq:LineElement}) is located at the radius $r=r_H$ defined by the largest positive root of the equation $B(r)=0$.\\

\begin{figure}[H]
	\centering
	\includegraphics[width=1 \linewidth]{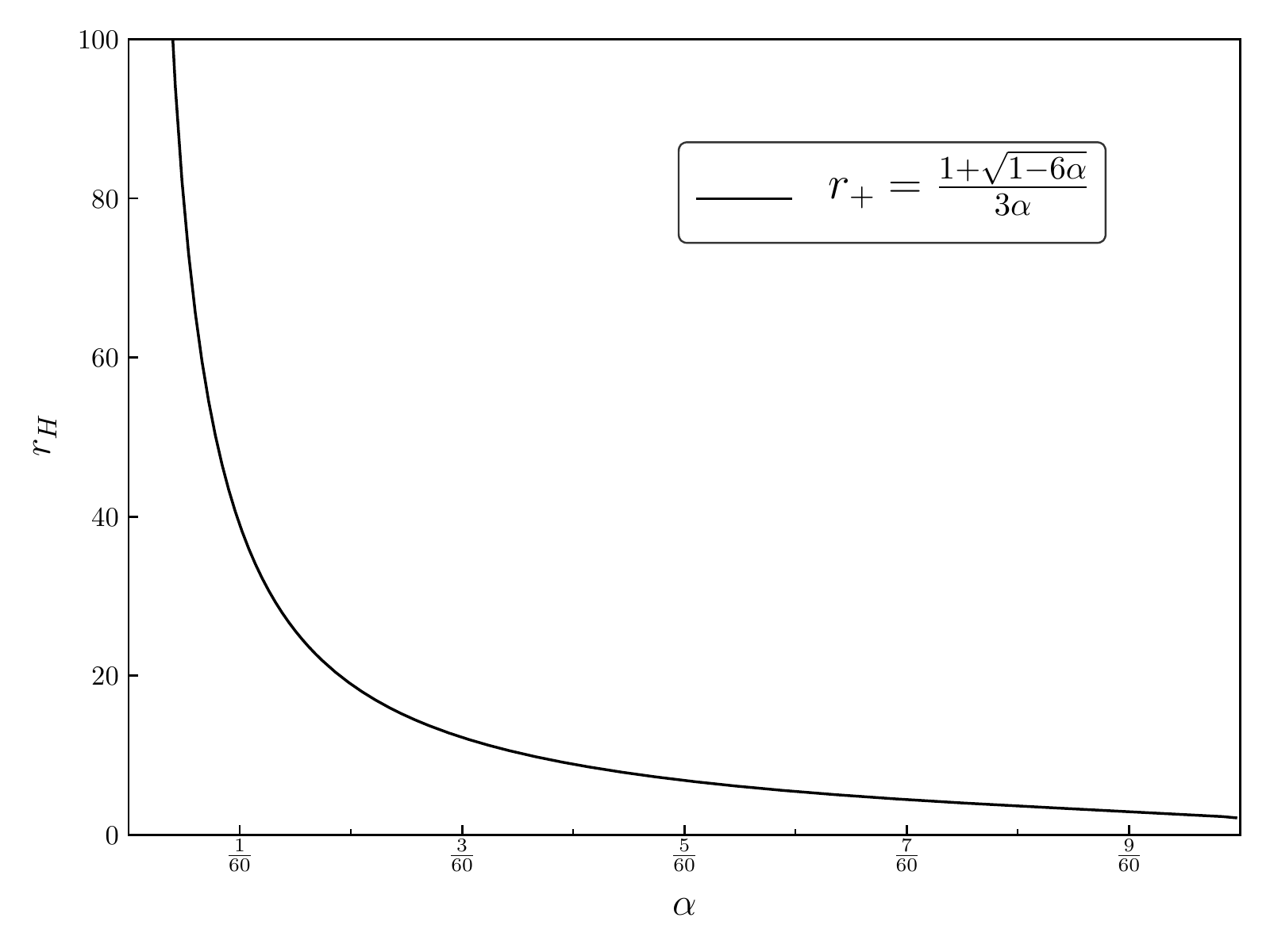}
	\caption{Radius of the event horizon as function of the dimensional parameter $\alpha$ for the asymptotically flat black hole.} 
	\label{fig:FlatBHHorizon}
\end{figure}

For the asymptotically flat black hole of Eq. (\ref{eq:FlatBH}), the horizon radius is 
\begin{equation}
    r_H(\alpha) = \frac{1}{3\alpha} \left[ 1 + \sqrt{1-6\alpha} \right]
\end{equation}
which differs from that reported in Eq. (26) in \cite{Nashed2019}. It is clear that the existence of the horizon implies that $0<\alpha\leq \frac{1}{6}$. The value $\alpha = \frac{1}{6}$ corresponds to an \textit{extremal} black hole with a degenerate horizon radius $r_H^{ext} = 2$ (see the complete behavior in Fig. \ref{fig:FlatBHHorizon}).\\

\subsection{The (Anti-)de Sitter Black Holes}
Considering a non-vanishing cosmological constant $\Lambda$, Nashed and Capozziello obtained a solution representing a charged, spherically symmetric black hole which asymptotically behaves as (anti-)de Sitter space-time. The line element has the same structure given in Eq. (\ref{eq:LineElement}) but this time
\begin{equation}
    B(r) = \frac{1}{2} - \frac{2\Lambda}{3}r^2 -\frac{1}{3\alpha r} + \frac{1}{3\alpha r^2}, \label{eq:AdSBH}
\end{equation}
while the gauge potential is given again by
\begin{equation}
    A^\mu (r) = \frac{1}{\sqrt{3\alpha} r} dt.
\end{equation}

This metric gives the Ricci's and Kretchamnn'a curvature invariants
\begin{widetext}
\begin{eqnarray}
    R &=& g^{\mu \nu} R_{\mu \nu} = \frac{1 + 8\Lambda r^2}{r^2} \label{eq:RicciScalar} \\
    K &=& R_{\alpha \beta \mu \nu} R^{\alpha \beta \mu \nu} = \frac{96 \Lambda^2 \alpha^2 r^8+ 24 \Lambda \alpha^2 r^6 + 9\alpha^2 r^4 + 12 \alpha r^3 - 12 (\alpha -1) r^2 - 48 r +56}{9 \alpha^2 r^8},
\end{eqnarray}
\end{widetext}
that show again that the point $r=0$ is an essential singularity and that $\alpha \neq 0$.\\

Concerning the event horizon, the condition $B(r_H) =0$ for the asymptotically (anti-)de Sitter black hole gives the horizon radius as the largest root of the polynomial
\begin{equation}
   4\Lambda \alpha r_H^4 - 3\alpha r_H^2 + 2r_H - 2 = 0.
\end{equation}

\begin{figure}[htb!]
	\centering
	\includegraphics[width=1 \linewidth]{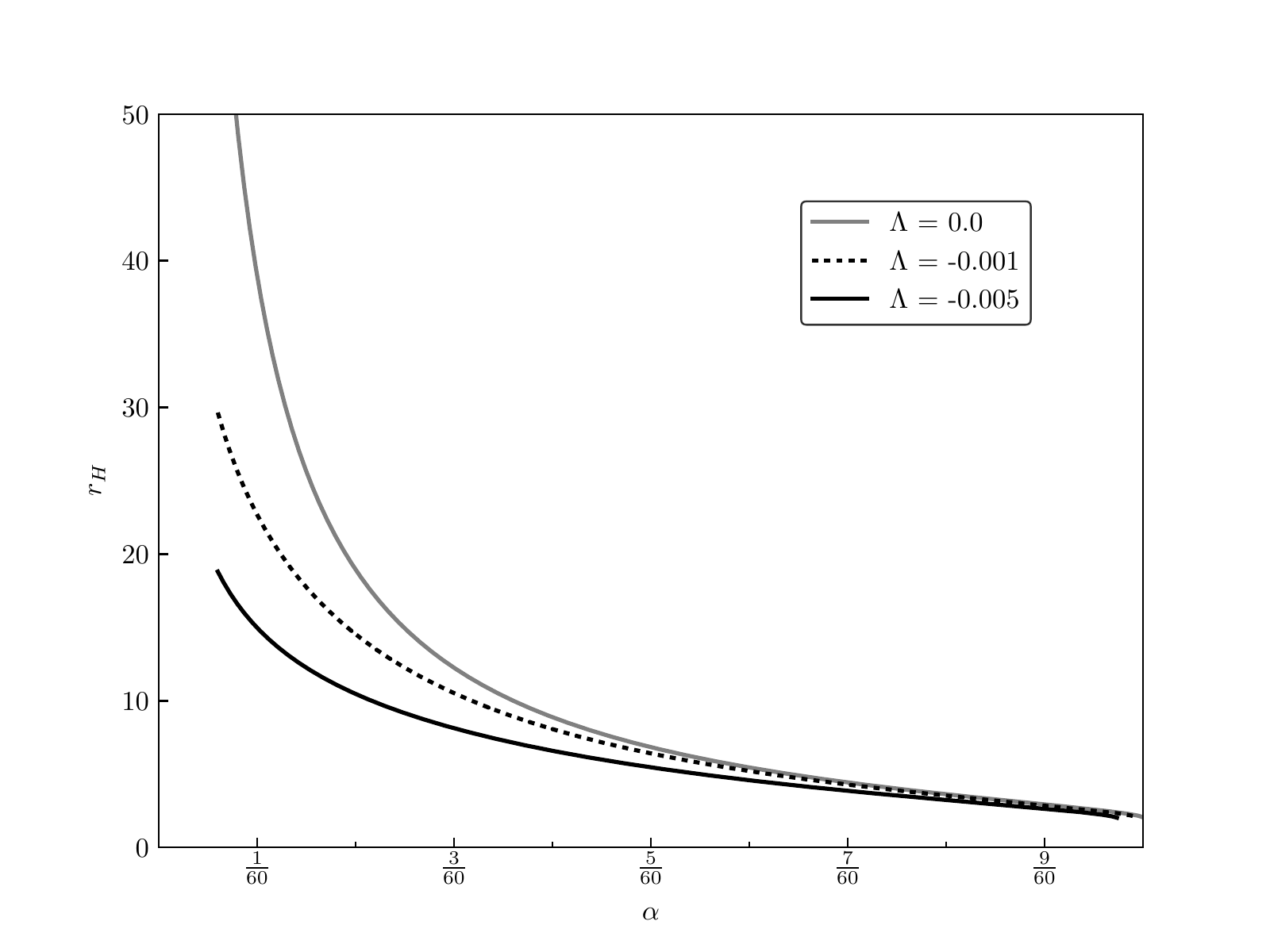}
	\caption{Radius of the event horizon as function of the dimensional parameter $\alpha$ for some values of the cosmological constant in the asymptotically (anti-)de Sitter black hole.} 
	\label{fig:AdSBHHorizon}
\end{figure}

This equation also differs from that reported in Eq. (26) in \cite{Nashed2019} by a sign in the last term. A numerical analysis let us plot the horizon radius as function of $\alpha$ as shown in Fig. \ref{fig:AdSBHHorizon}.

\section{Black Hole Thermodynamics}
In this section we will study the black hole thermodynamic properties such as Hawking temperature, entropy, quasi-local energy and Gibbs free energy. As usual, the Hawking temperature for these spherically symmetric black holes is given by the expression
\begin{equation}
    T_H = \frac{1}{4\pi} \left. \frac{dB}{dr} \right|_{r=r_H}
\end{equation}
while the Bekenstein-Hawking entropy and the quasi-local energy in the $f(R)$ model of gravity are \cite{Cognola2011, Sheykhi2012, Zheng2018, Nashed2018} 
\begin{equation}
    S = \frac{1}{4} A_H f'\left(R(r_H)\right)
\end{equation}
and
\begin{eqnarray}
    E_H = \frac{1}{4} \int \left[ 2f'\left( R(r_H) \right) + r_H^2 f\left( R(r_H)\right) \right. \notag \\
    \left. -r_H^2 R(r_H) f'\left(R(r_H) \right) \right] dr_H,
\end{eqnarray}
where $A_H = 4 \pi r_H^2$ is the area of the event horizon.

\subsection{Asymptotically Flat Black Hole}
 The temperature associated with the horizon of the asymptotically flat black hole, see Eq. (\ref{eq:FlatBH}), gives the function
 
\begin{equation}
    T_H(\alpha) = \frac{3\alpha (1+ \sqrt{1-6\alpha} - 6\alpha)}{4\pi (1+\sqrt{1-6\alpha})^3}.
\end{equation}

\begin{figure}[htb!]
	\centering
	\includegraphics[width=1 \linewidth]{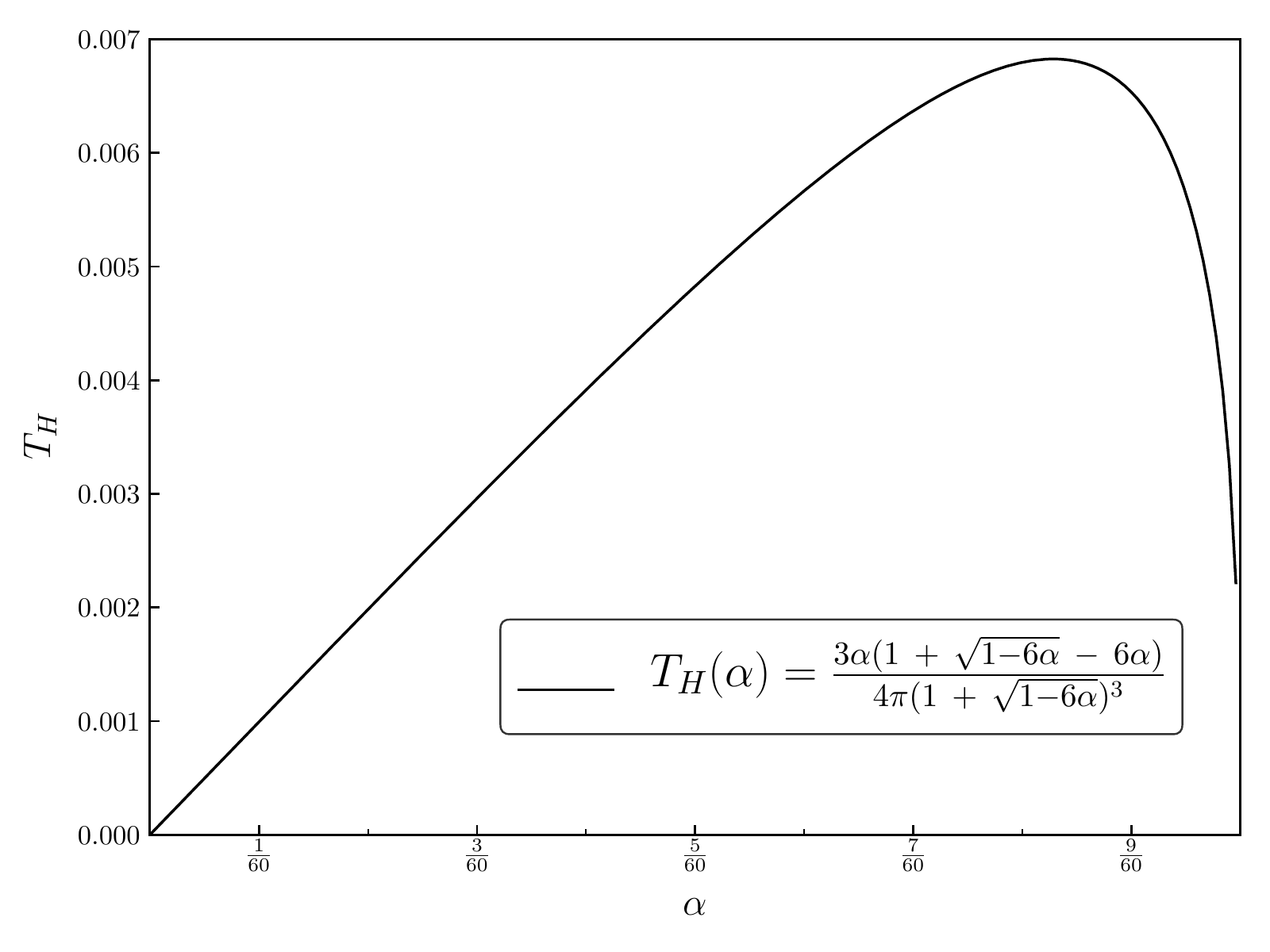}
	\caption{Hawking temperature as function of the dimensional parameter $\alpha$ for the asymptotically flat black hole.} 
	\label{fig:FlatBHTemperature}
\end{figure}

The complete behavior of this temperature as function of the parameter $\alpha$ is significantly different from that reported in \cite{Nashed2019} and can be seen in Fig. \ref{fig:FlatBHTemperature}. The value $\alpha = \frac{1}{6}$, corresponding to the extremal black hole, gives a zero temperature, just as in black holes such as Reissner-Nördstrom or Kerr solutions in general relativity.  Hence, the process of evaporation of these black holes will terminate in a frozen remnant with a final horizon radius $r_H^{ext} = 2$. It is also interesting to note that the temperature function presents a maximum of $T_H^{max} \approx 0.00682667 $ at the parameter $\alpha_m = \frac{1}{3}\left( \sqrt{2} -1\right) \approx 0.138071$. This behavior is a unique characteristic of these black holes, with no similar situations in general relativity solutions.\\

\begin{figure}[htb!]
	\centering
	\includegraphics[width=1 \linewidth]{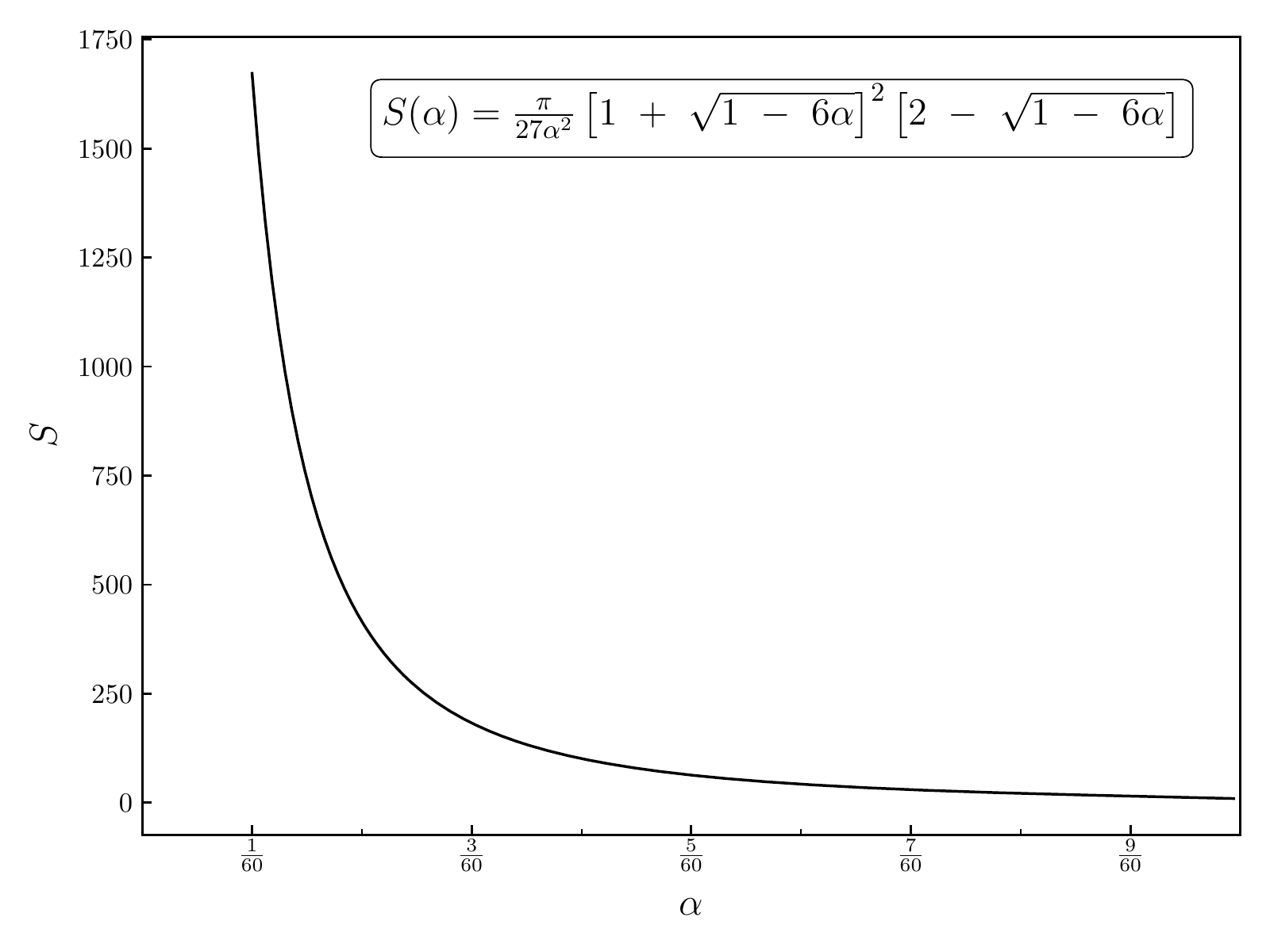}
	\caption{Bekenstein-Hawking entropy as function of the parameter $\alpha$ for the asymptotically flat black hole.} 
	\label{fig:FlatBHEntropy}
\end{figure}

On the other hand, the entropy for the asymptotically flat black hole is the function
\begin{equation}
    S(\alpha) = \frac{\pi}{27\alpha^2}\left[1+\sqrt{1-6\alpha} \right]^2 \left[2-\sqrt{1-6\alpha} \right],
\end{equation}
from which it is clear that $S(\alpha)$  is a smooth function and always positive in the range $0\leq \alpha <\frac{1}{6}$, as shown in Fig. \ref{fig:FlatBHEntropy}. Therefore, there is no such a phase transition as the one reported in the analysis of Nashed and Capozziello \cite{Nashed2019}. 

To complete the thermodynamic analysis, we calculate the quasi-local energy for the asymptotically flat black hole, obtaining
\begin{equation}
    E(\alpha) = \frac{1+\sqrt{1-6\alpha} + 3\alpha}{12\alpha}
\end{equation}
and then, the Gibbs free energy gives
\begin{equation}
    G(\alpha) = \frac{5}{36 \alpha} - \frac{(1 - \sqrt{1-6\alpha})}{12(1 + \sqrt{1-6\alpha})}.
\end{equation}

\begin{figure}[htb!]
	\centering
	\includegraphics[width=1 \linewidth]{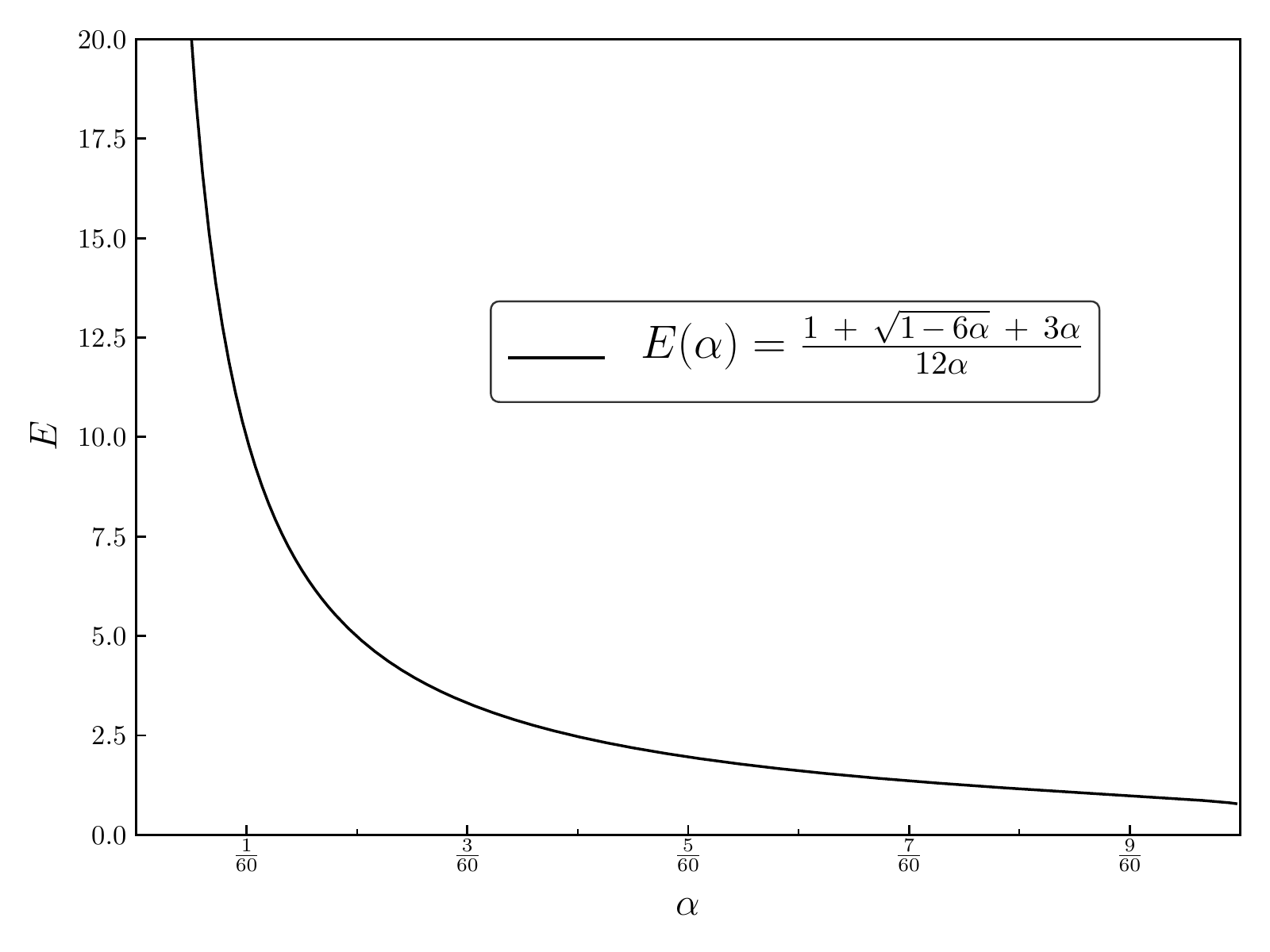}
	\caption{Quasi-local energy as function of the parameter $\alpha$ for the asymptotically flat black hole.} 
	\label{fig:FlatBHCuasiLocalEnergy}
\end{figure}

\begin{figure}[htb!]
	\centering
	\includegraphics[width=1 \linewidth]{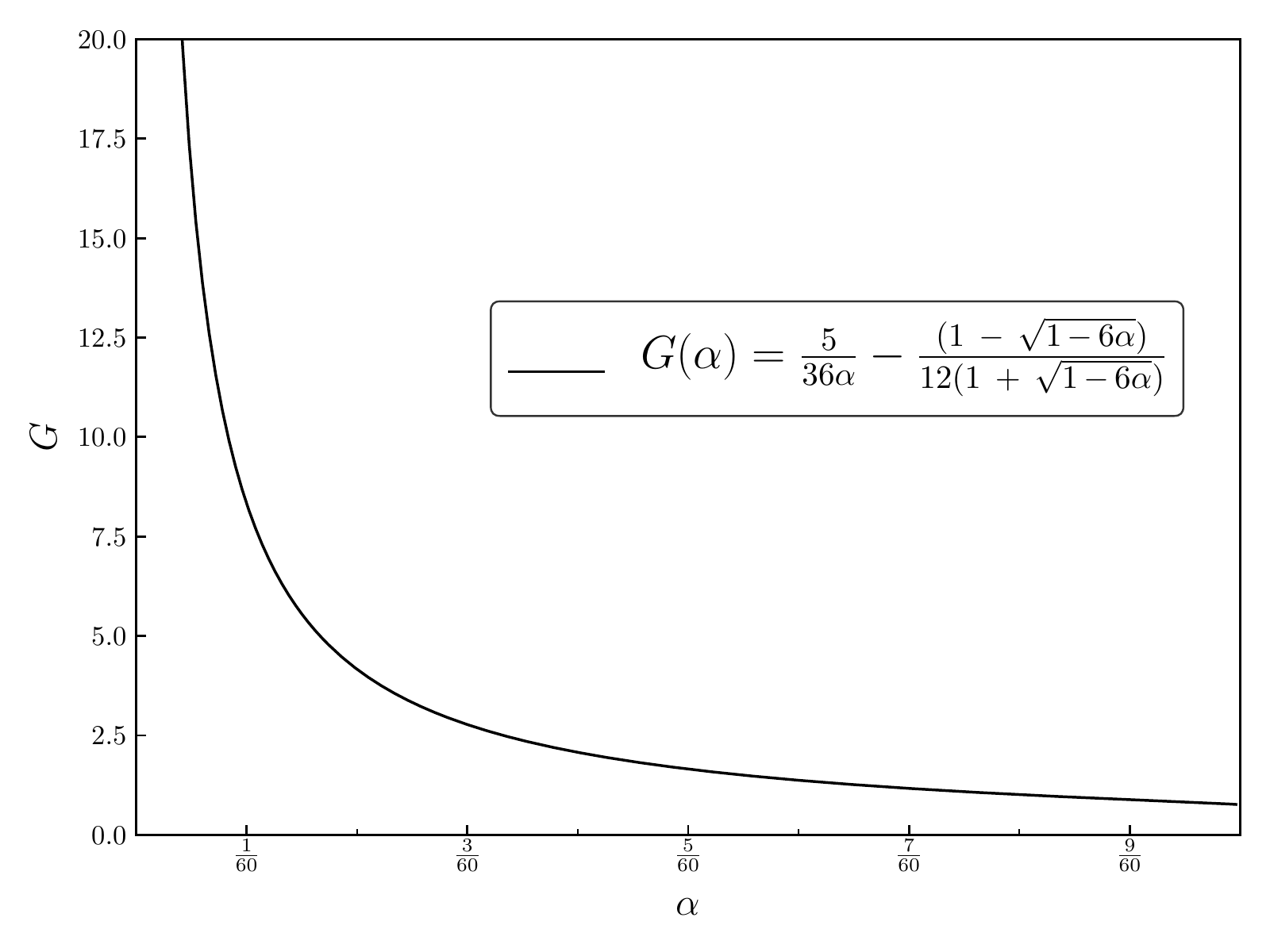}
	\caption{Gibbs free energy as function of the parameter $\alpha$ for the asymptotically flat black hole.} 
	\label{fig:FlatBHGibbs}
\end{figure}

The behavior of these function is shown in Figures \ref{fig:FlatBHCuasiLocalEnergy} and \ref{fig:FlatBHGibbs}. Both of them have a positive value and smooth behavior in the allowed range of the parameter $\alpha$. This is a clear indication that no phase transitions occur for the asymptotically flat black hole \cite{Altamirano2014}.

\subsection{Asymptotically (Anti-)de Sitter Black Hole}

\begin{figure}[htb!]
	\centering
	\includegraphics[width=1 \linewidth]{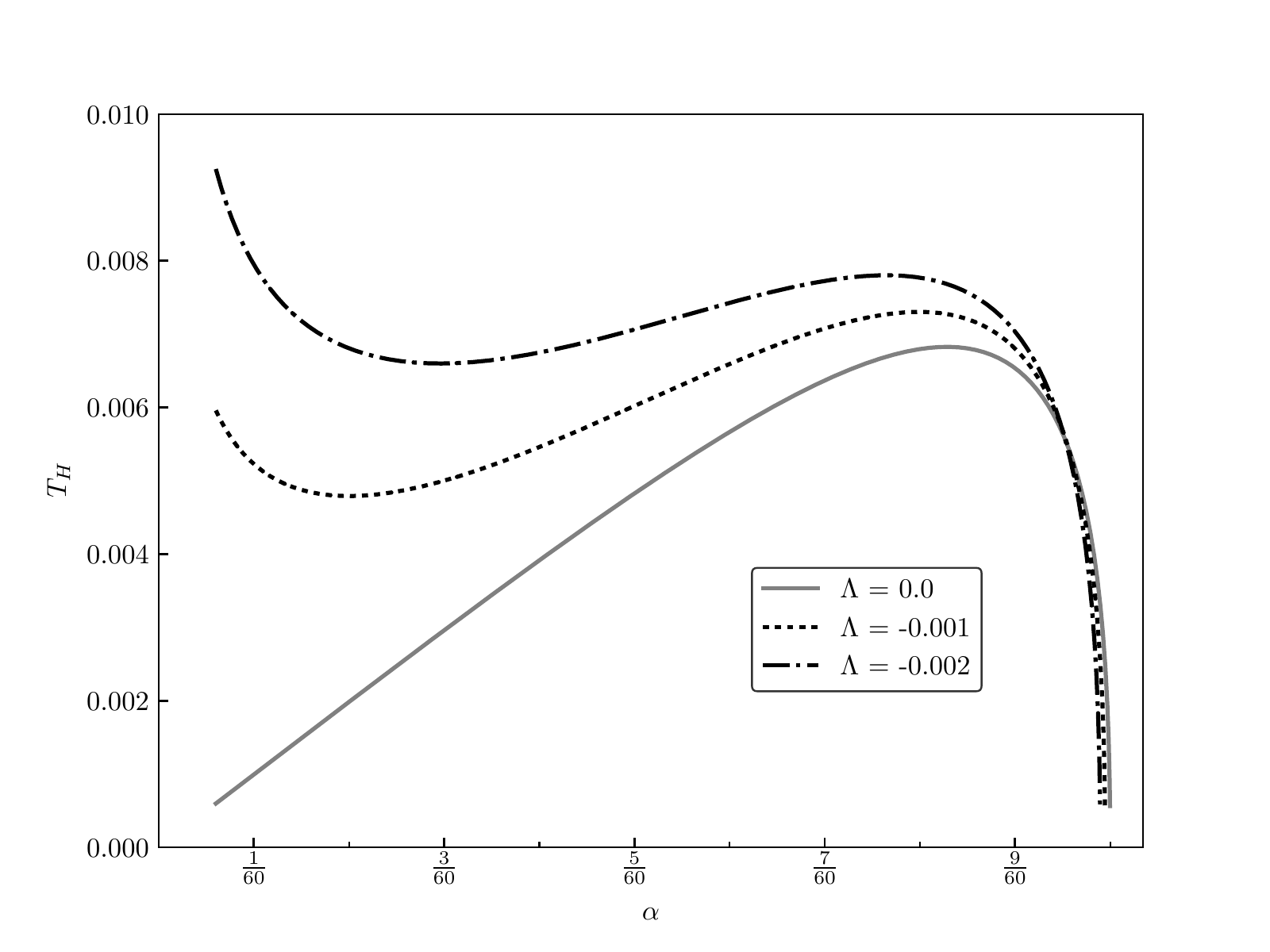}
	\caption{Hawking temperature $T_{H}$ as function of the dimensional parameter $\alpha$ for some values of the cosmological constant in the asymptotically (anti-)de Sitter black hole.} 
	\label{fig:AdSBHTemperature}
\end{figure}

\begin{figure}[htb!]
	\centering
	\includegraphics[width=1 \linewidth]{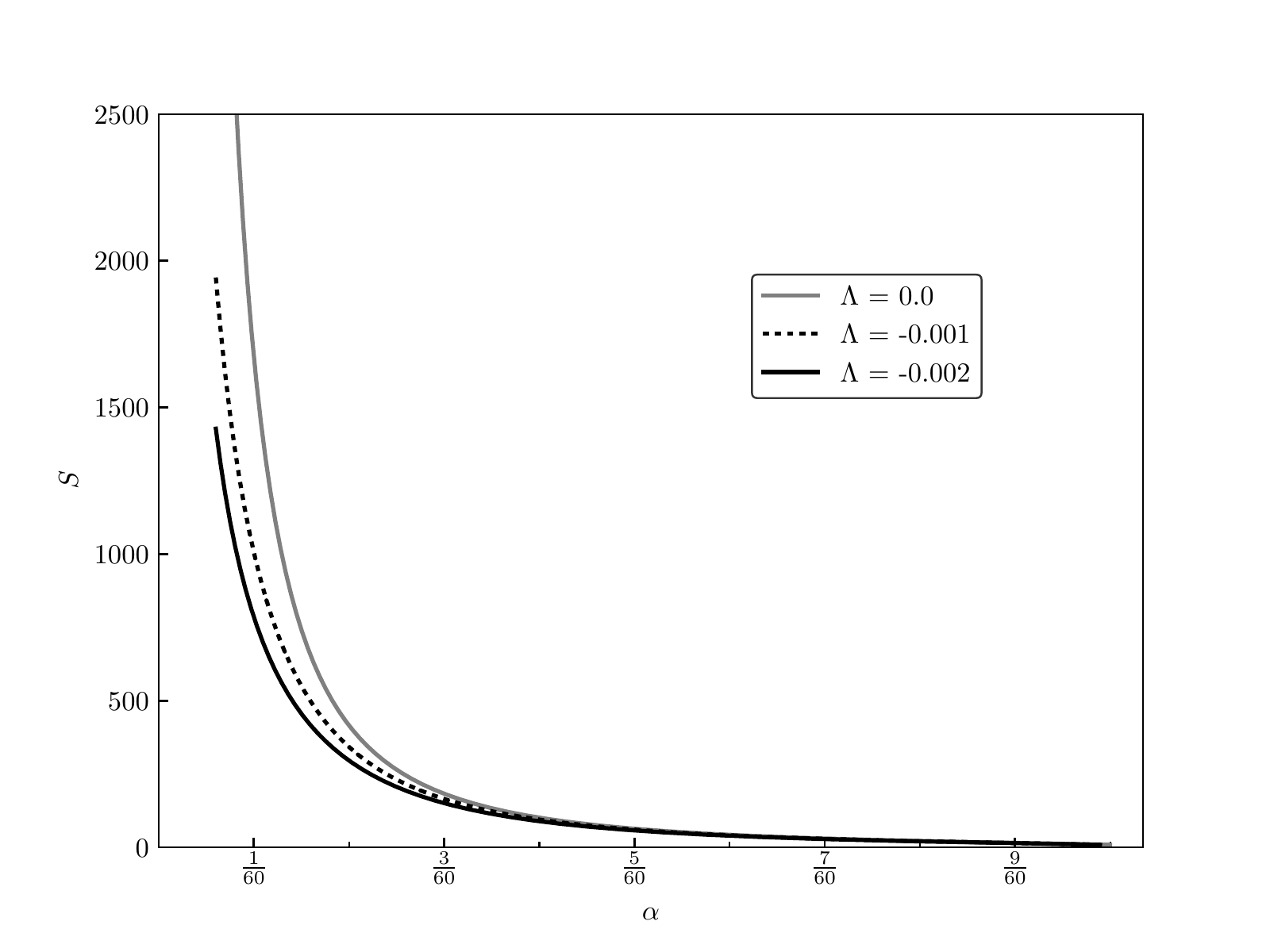}
	\caption{Bekenstein-Hawking entropy $S$ as function of the dimensional parameter $\alpha$ for some values of the cosmological constant in the asymptotically (anti-)de Sitter black hole.} 
	\label{fig:AdSBHEntropy}
\end{figure}

Considering now the contribution of the cosmological constant in Eq. (\ref{eq:AdSBH}), the Hawking temperature and entropy are given by
\begin{equation}
    T_H(\alpha) = \frac{r_H - 2 -4\alpha \Lambda r_H^4}{12\pi \alpha r_H^3}
\end{equation}
and
\begin{equation}
    S = \pi r_H^2 \left( 1- \alpha r_H \right),
\end{equation}
where the value of the horizon radius $r_H$ must be calculated numerically. The behavior of these functions for some particular values of $\Lambda$ is shown in Figures \ref{fig:AdSBHTemperature} and \ref{fig:AdSBHEntropy}. Concerning the temperature, note that the cosmological constant produces an interesting behavior for small $\alpha$, making $T$ to diverge. Hence, the temperature function has a minimum and a maximum, depending on the value of $\Lambda$. However, there is a value of $\alpha$ for which the black hole freezes ($T\rightarrow 0$). \\
On the other hand, the entropy function does not change significantly with the introduction of the cosmological constant, it has positive values and a smooth behavior in the allowed range of $\alpha$.

The quasi-local energy for the asymptotically (anti-)de Sitter black hole gives the function
\begin{equation}
    E(r_H) = \frac{r_H}{2} - \frac{3\alpha}{8}r_H^2 + \frac{\Lambda \alpha}{2}r_H^4 
\end{equation}
and therefore, the Gibbs energy takes the form
\begin{eqnarray}
   G (r_H) &=& \frac{1}{2} \alpha  \Lambda  r^4-\frac{3 \alpha  r^2}{8}+\frac{r}{2} \notag \\
   & & -\frac{(1-\alpha  r) \left(-4 \alpha  \Lambda  r^4+r-2\right)}{12 \alpha  r}.
\end{eqnarray}

\begin{figure}[htb!]
	\centering
	\includegraphics[width=1 \linewidth]{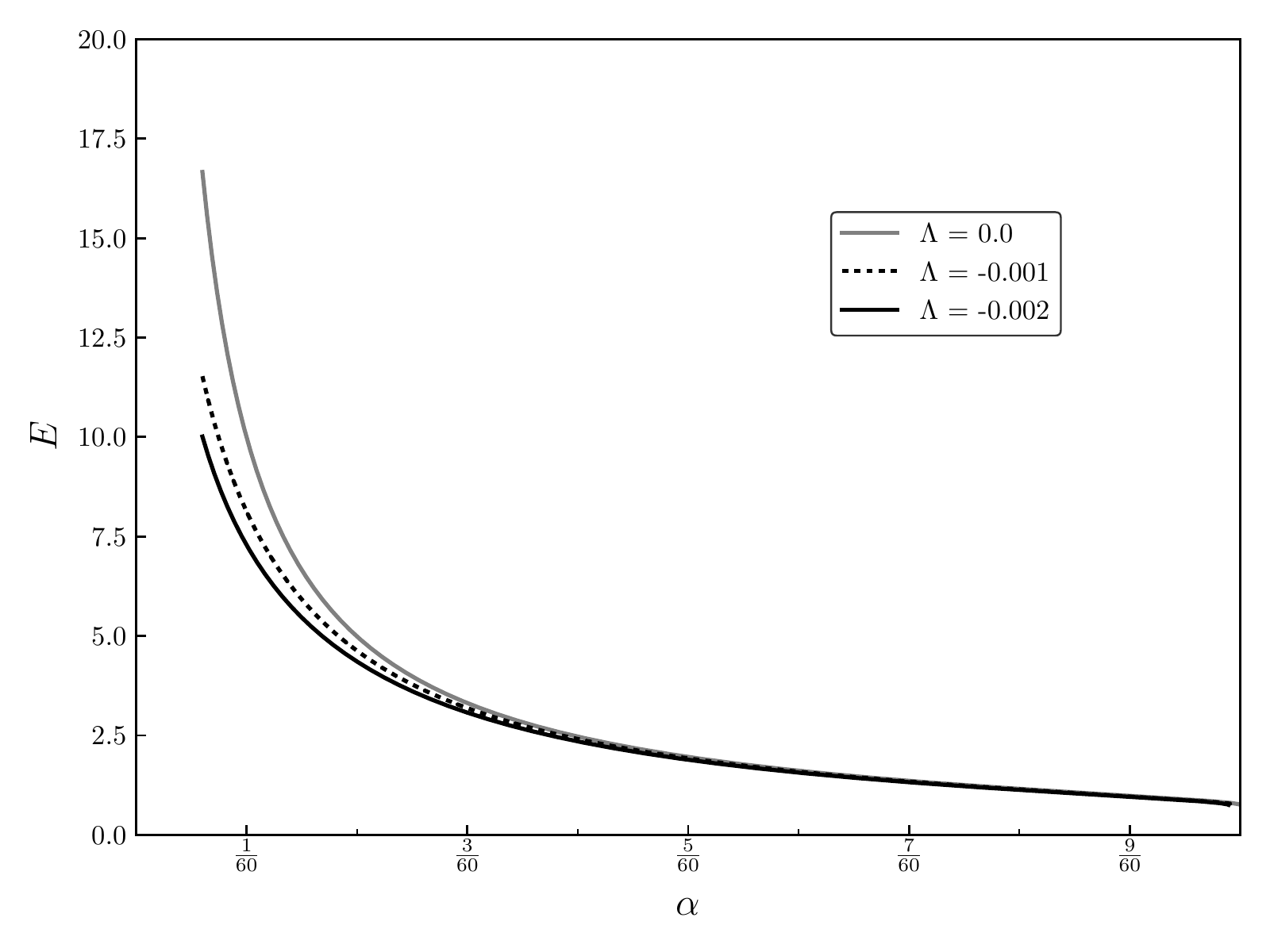}
	\caption{Quasi-local energy as function of the parameter $\alpha$ for the asymptotically (anti-)de Sitter black hole.} 
	\label{fig:AdSBHQuasiLocalEnergy}
\end{figure}

\begin{figure}[htb!]
	\centering
	\includegraphics[width=1 \linewidth]{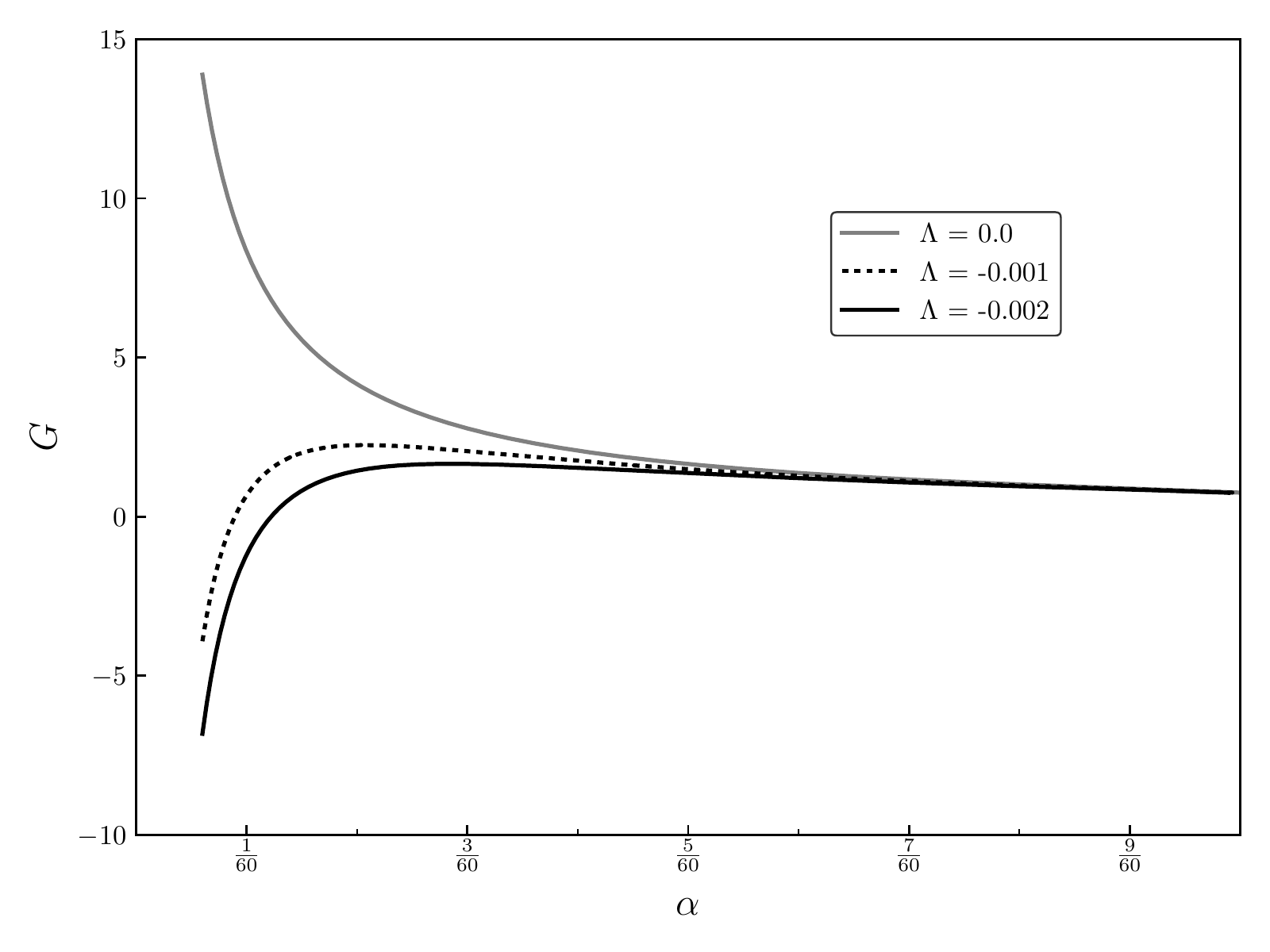}
	\caption{Gibs free energy as function of the parameter $\alpha$ for the asymptotically (anti-)de Sitter black hole.} 
	\label{fig:AdSBHGibbs}
\end{figure}

In Figures \ref{fig:AdSBHQuasiLocalEnergy} and \ref{fig:AdSBHGibbs} we plot these functions for some particular values of the cosmological constant. Both of them have a smooth behavior with no indication of phase transitions for the asymptotically (A)dS black holes.

\section{Geodesics Stability}

\subsection{Circular Geodesics}
The study of the geodesics describing a test particle moving along a trajectory $x^{\alpha}(\lambda)$ with 4-velocity $\dot{x}^{\alpha}=dx^{\alpha}/d\lambda$ around one of the described black holes, gives the conservation of the specific energy, $E$, and the specific axial component of the angular momentum, $l$,
 \begin{equation}
    E = B(r) \dot{t} \hspace{0.5cm} , \hspace{0.5cm} l=r^{2}\dot{\phi} \ \ ,
 \end{equation}
where the dot represents derivative with respect to the particle's proper time.
From the conservation of the rest-mass $g_{\mu \nu}\dot{x}^{\mu}\dot{x}^{\nu}=1$, we obtain the equation of motion for the radial coordinate in the equatorial plane ($\theta = \pi /2$),
\begin{equation}
    \dot{r}^{2}=E^{2}-V^{2}_{\textrm{eff}}(r)
    \label{eq:radialequation}
\end{equation}
where the effective potential function is 
\begin{equation}
    V_{\textrm{eff}}^{2}(r) = B(r)\left(1+\frac{l^{2}}{r^{2}}\right)  \ .
    \label{eq:Veff}
\end{equation}

\begin{figure}[H]
	\centering
	\includegraphics[width=1.0 \linewidth]{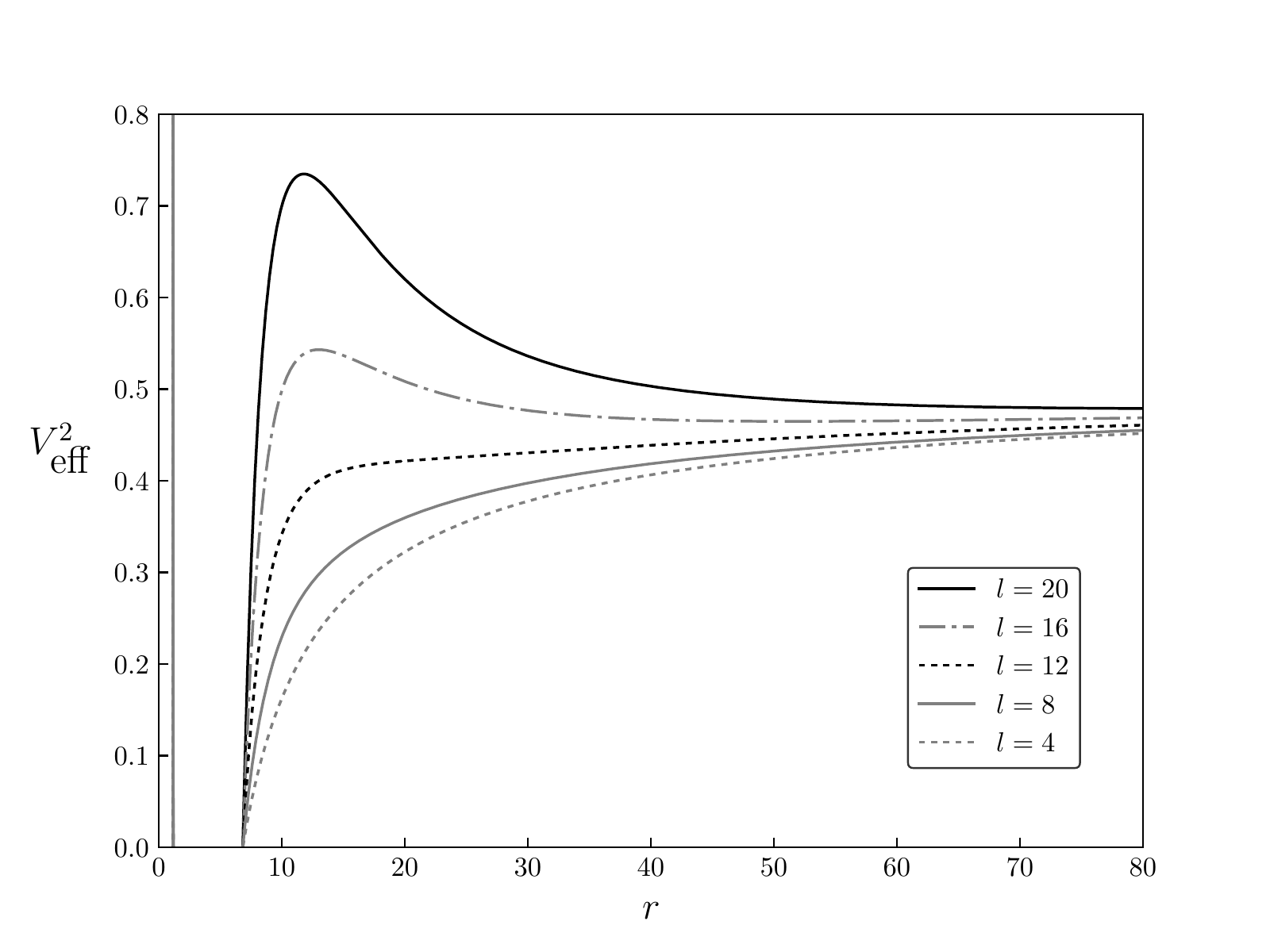}
	\caption{Effective potential as function of the radial coordinate for different values of angular momentum with the particular choice $\alpha = 1/12$ and $\Lambda = 0$.} 
	\label{fig:Vef}
\end{figure}

The behavior of the effective potential for the asymptotically flat black hole (\ref{eq:FlatBH}) is shown in Figure \ref{fig:Vef}. Note that the form of the curve indicates the existence of bound and unbound orbits, as well as the possibility of circular trajectories. These are defined by the conditions
\begin{equation}
    \theta = \frac{\pi}{2},\hspace{1.5em} 
    \frac{d\theta}{d\lambda} = 0, \hspace{1.5em}
    \frac{dr}{d\lambda} = 0. \label{eq:circularOrbit1} 
\end{equation}
together with the equations of motion for $t$ and $\phi$ which, using the conservation laws, are
\begin{eqnarray}
    \left( \frac{d\phi}{d\lambda} \right)^2 &=& \frac{\partial_r B}{r (2B-r \partial_r B)} \notag \\
     \left( \frac{dt}{d\lambda} \right)^2 &=& \frac{2}{(2B-r \partial_r B)}. \label{eq:circularOrbits2}
\end{eqnarray}

From these equations, it is possible to write the angular velocity $\Omega = \dot{\phi}/\dot{t}$ as 
\begin{equation}
    \Omega = \sqrt{\dfrac{\partial_{r}B(r)}{2r}}.
\end{equation}

\begin{figure}[h!]
	\centering
	\includegraphics[width=1 \linewidth]{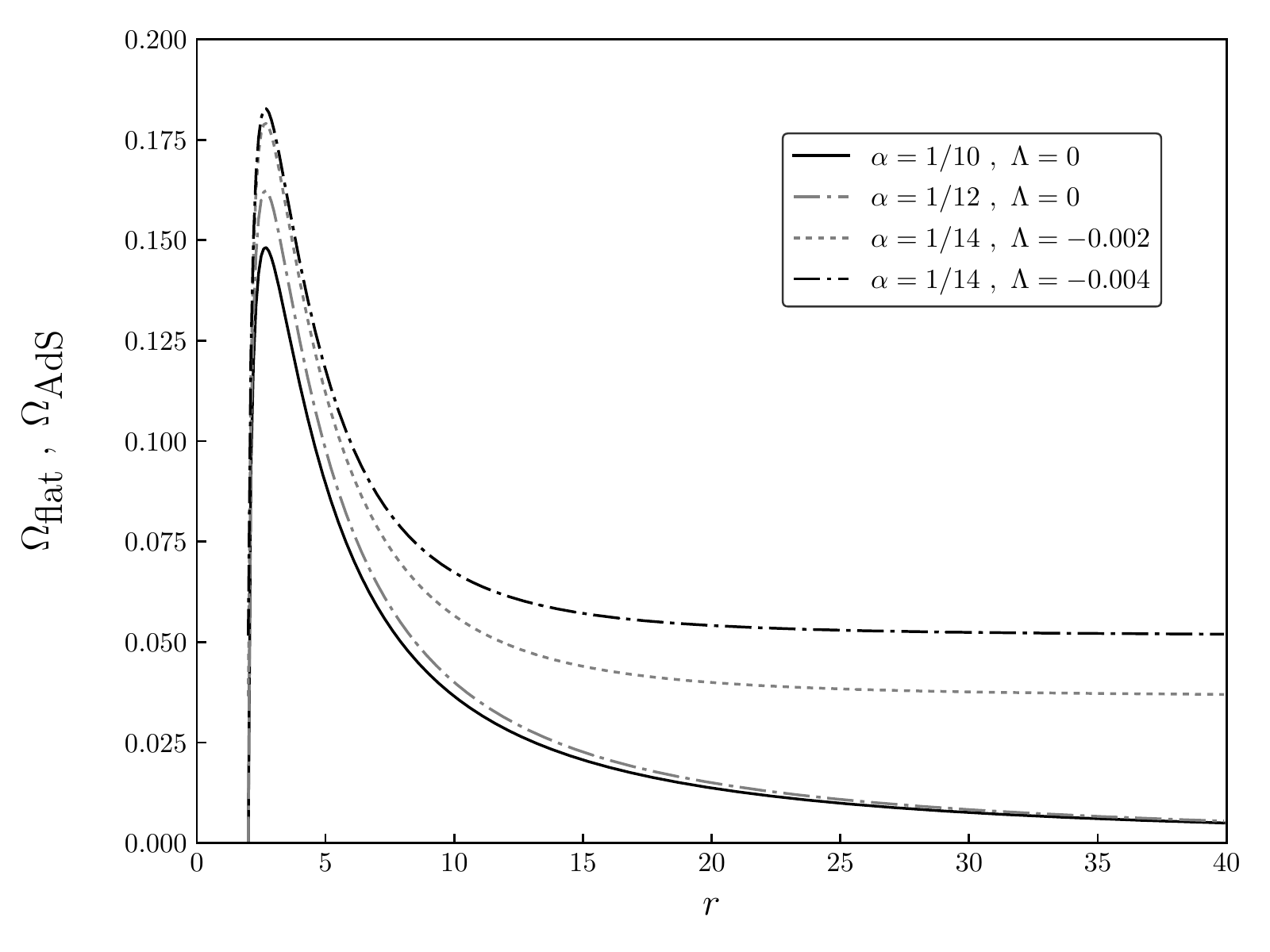}
	\caption{Angular velocity of a particle moving around the $f(R)$ black holes as function of the radius of the circular orbit.} 
	\label{fig:AngularVelocity}
\end{figure}

 For the asymptotically flat solution in equation (\ref{eq:FlatBH}) and for the (anti-)de Sitter solution in (\ref{eq:AdSBH}) we get the angular velocities
 \begin{equation}
     \Omega_{\mathrm{flat}} = \sqrt{\dfrac{r-2}{6\alpha r^{4}}} \hspace{0.5cm}, \hspace{0.5cm}  \Omega_{\mathrm{AdS}} = \sqrt{\dfrac{r-2-4\Lambda \alpha r^{4}}{6\alpha r^{4}}} \ \ ,
 \end{equation}
which correspond to Kepler's third law for particles moving around these two black holes. The behavior of these angular velocities is shown in Figure \ref{fig:AngularVelocity}.

Due to the characteristics of the black hole metrics, it is interesting to study the Innermost Stable Circular Orbit (ISCO), which is determined by the conditions
\begin{equation}
      \frac{d V_{\textrm{eff}}^{2}}{dr} \Big|_{r = r_{_{\textrm{ISCO}}}}= 0  \hspace{0.5cm} \text{ and }\hspace{0.5cm}
     \frac{d^{2}V_{\textrm{eff}}^{2}}{dr^{2}} \Big|_{r = r_{_{\textrm{ISCO}}}}= 0. 
\end{equation}

These equations combine to give the radius of the ISCO through the relation
\begin{equation}
    3B(r) - 2rB'(r) + \frac{rB''(r)B}{B'} = 0.
    \label{eq:r_ISCOCondition1}
\end{equation}

In Fig. \ref{fig:Vef} we show how the effective potential changes for different values of angular momentum with the particular values $\alpha=\frac{1}{12}$ and $\Lambda = 0$. There is an ISCO for $l_{ISCO}\approx12.7848$ at $r_{ISCO}\approx20.6658$ and we notice that as $l$ increases from this value, there are both stable and unstable circular orbits.

\begin{figure}[htb!]
	\centering
	\includegraphics[width=1 \linewidth]{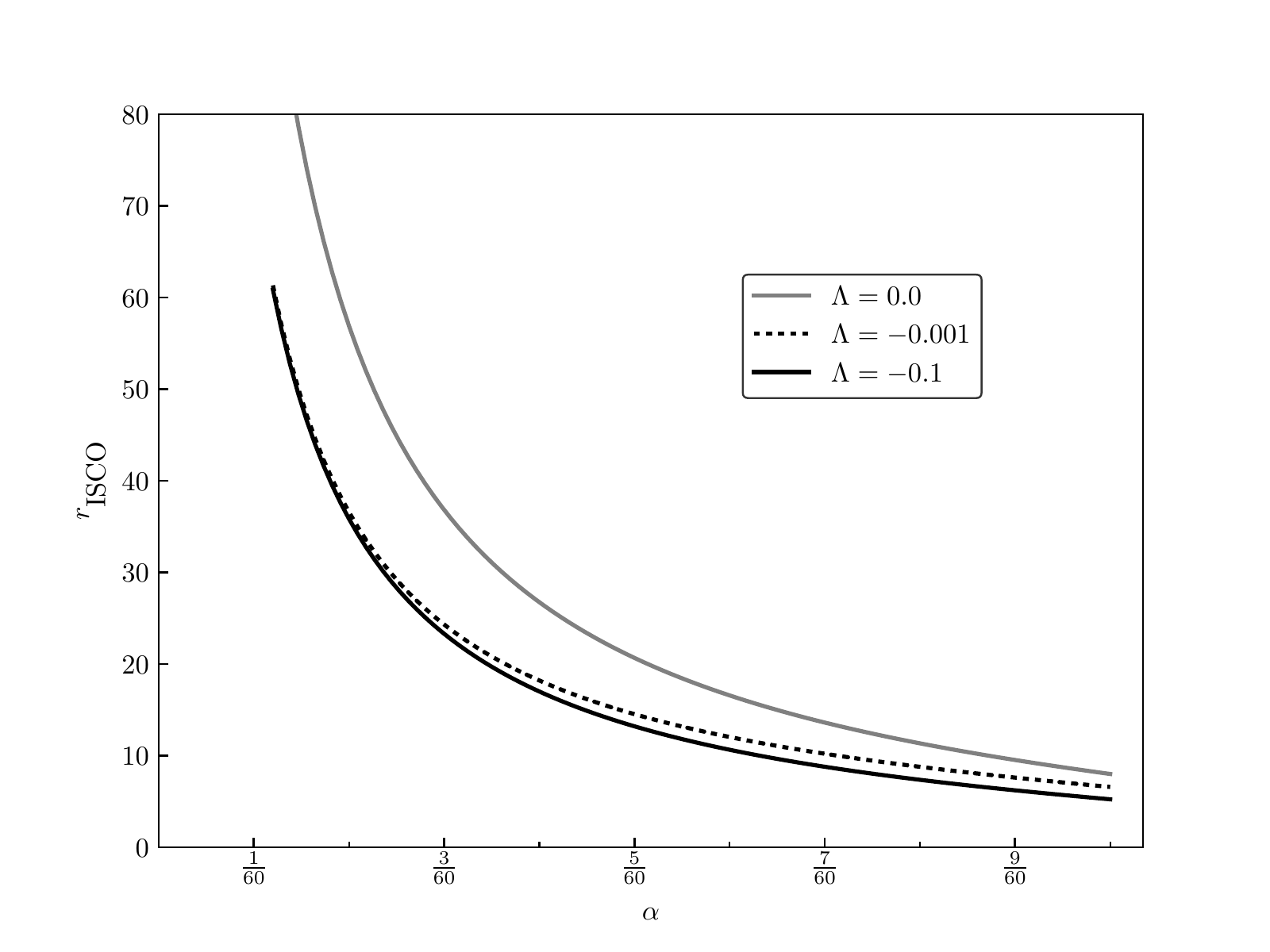}
	\caption{Radius of the ISCO as function of the parameter $\alpha$ for some particular values of the cosmological constant.} 
	\label{fig:BHrISCO}
\end{figure}

By numerically solving equation (\ref{eq:r_ISCOCondition1}) using the function $B(r)$ in equation (\ref{eq:AdSBH}) we obtain the radius of the ISCO as function of the parameter $\alpha$ for some values of the cosmological constant, as shown in Figure \ref{fig:BHrISCO}.

\subsection{Geodesic Deviation Equations}

In order to study the stability of the circular orbits we will consider the geodesic deviation equation \cite{MaggioreGW},
\begin{equation}
    \frac{d^2 \xi^\alpha}{d\lambda^2} + 2 \Gamma ^\alpha_{\mu \nu} \frac{dx^\mu}{d\lambda} \frac{d\xi^\nu}{d\lambda} +  \frac{dx^\mu}{d\lambda} \frac{dx^\nu}{d\lambda}\xi^\beta \partial_\beta \Gamma ^\alpha_{\mu \nu}  = 0,
\end{equation}
where $\xi^\alpha$ is the deviation 4-vector. By replacing the spherically symmetric line element (\ref{eq:LineElement}), we obtain the general set of  equations
\begin{widetext}
\begin{eqnarray}
    & &\frac{d^2 \xi^0}{d\lambda^2} + \frac{B'}{B} \frac{dt}{d\lambda} \frac{d\xi^1}{d\lambda} + \frac{B'}{B} \frac{dr}{d\lambda} \frac{d\xi^0}{d\lambda} +\frac{dt}{d\lambda} \frac{dr}{d\lambda} \xi^\beta \partial_\beta \left(\frac{B'}{B} \right)=0 \notag \\
   & & \frac{d^2 \xi^1}{d\lambda^2} + BB' \frac{dt}{d\lambda} \frac{d\xi^0}{d\lambda} - \frac{B'}{B} \frac{dr}{d\lambda} \frac{d\xi^1}{d\lambda} - 2rB \frac{d\theta}{d\lambda} \frac{d\xi^2}{d\lambda} - 2rB\sin^2 \theta \frac{d\phi}{d\lambda} \frac{d\xi^3}{d\lambda} 
    + \frac{1}{2} \left(\frac{dt}{d\lambda}\right)^2  \xi^\beta \partial_\beta \left( BB' \right) \notag \\
      & &-\frac{1}{2} \left(\frac{dr}{d\lambda}\right)^2  \xi^\beta \partial_\beta \left( \frac{B'}{B} \right) -  \left(\frac{d\theta}{d\lambda}\right)^2  \xi^\beta \partial_\beta \left( rB \right) -  \left(\frac{d\phi}{d\lambda}\right)^2  \xi^\beta \partial_\beta \left( rB \sin^2 \theta \right) = 0 \notag \\
    & &\frac{d^2 \xi^2}{d\lambda^2} + \frac{2}{r} \frac{dr}{d\lambda} \frac{d\xi^2}{d\lambda} + \frac{2}{r} \frac{d\theta}{d\lambda} \frac{d\xi^1}{d\lambda}
    - 2 \sin \theta \cos \theta \frac{d\phi}{d\lambda} \frac{d\xi^3}{d\lambda}
    + 2 \frac{dr}{d\lambda} \frac{d\theta}{d\lambda}  \xi^\beta \partial_\beta \left( \frac{1}{r} \right)
    -  \left(\frac{d\phi}{d\lambda}\right)^2  \xi^\beta \partial_\beta \left( \sin \theta \cos \theta\right) = 0 \notag \\
   & & \frac{d^2 \xi^3}{d\lambda^2} + \frac{2}{r} \frac{dr}{d\lambda} \frac{d\xi^3}{d\lambda} + \frac{2}{r} \frac{d\phi}{d\lambda} \frac{d\xi^1}{d\lambda} + \frac{2\cos \theta}{\sin \theta} \frac{d\theta}{d\lambda} \frac{d\xi^3}{d\lambda} + \frac{2\cos \theta}{\sin \theta} \frac{d\phi}{d\lambda} \frac{d\xi^2}{d\lambda} + 2 \frac{dr}{d\lambda} \frac{d\phi}{d\lambda}  \xi^\beta \partial_\beta \left( \frac{1}{r} \right) + 2 \frac{d\theta}{d\lambda} \frac{d\phi}{d\lambda}  \xi^\beta \partial_\beta \left( \frac{\cos \theta}{\sin \theta} \right) = 0. 
\end{eqnarray}
\end{widetext}

For a circular orbit satisfying conditions (\ref{eq:circularOrbit1}), the geodesic deviation equations reduce to
\begin{eqnarray}
    & &\frac{d^2 \xi^0}{d\lambda^2} + \frac{B'}{B} \frac{dt}{d\lambda} \frac{d\xi^1}{d\lambda}=0 \notag \\
    & &\frac{d^2 \xi^1}{d\lambda^2} + BB' \frac{dt}{d\lambda} \frac{d\xi^0}{d\lambda} - 2rB \frac{d\phi}{d\lambda} \frac{d\xi^3}{d\lambda}   \notag \\
    & &+ \left[\frac{1}{2} \left(\frac{dt}{d\lambda}\right)^2  \left( B'^2 + B B'' \right)
     -  \left(\frac{d\phi}{d\lambda}\right)^2   \left( B + rB' \right) \right] \xi^1 = 0 \notag \\
   & & \frac{d^2 \xi^2}{d\lambda^2} 
    +  \left(\frac{d\phi}{d\lambda}\right)^2  \xi^2  = 0 \notag \\
    & &\frac{d^2 \xi^3}{d\lambda^2} + \frac{2}{r} \frac{d\phi}{d\lambda} \frac{d\xi^1}{d\lambda}  = 0, 
\end{eqnarray}
and using the equations of motion (\ref{eq:circularOrbits2}), they become
\begin{eqnarray}
    & &\frac{d^2 \xi^0}{d\phi^2} + \frac{B'}{B} \frac{dt}{d\phi} \frac{d\xi^1}{d\phi}=0 \notag \\
    & &\frac{d^2 \xi^1}{d\phi^2} + BB' \frac{dt}{d\phi} \frac{d\xi^0}{d\phi} - 2rB  \frac{d\xi^3}{d\phi}  \notag \\
    & &+ \left[\frac{1}{2}\left(\frac{dt}{d\phi}\right)^2  \left( B'^2 + B B'' \right)
     - \left( B + rB' \right) \right] \xi^1 = 0 \notag \\
    & &\frac{d^2 \xi^2}{d\phi^2} 
    +  \xi^2  = 0 \notag \\
    & &\frac{d^2 \xi^3}{d\phi^2} + \frac{2}{r}  \frac{d\xi^1}{d\phi}  = 0. \label{eq:GeodesicDeviationCircular}
\end{eqnarray}

It is clear that the equation for $\xi^2$ shows stability because its solution $\xi^2 = \zeta^2 e^{i\phi}$ indicates that a tests particle initially moving in the equatorial plane will perform harmonic motion around it under perturbations. 

For the remaining variables, we suppose solutions with the form
\begin{eqnarray}
    \xi^0 &=& \zeta^0 e^{i \omega \phi} \notag \\
    \xi^1 &=& \zeta^1 e^{i \omega \phi} \notag \\
    \xi^3 &=& \zeta^3 e^{i \omega \phi}.
\end{eqnarray}

Replacing in (\ref{eq:GeodesicDeviationCircular}) and requiring stability for the circular motion, we obtain the condition
\begin{equation}
    \omega^2 = 3B - 2rB' + \frac{rBB''}{B'} > 0.
    \label{eq:omega}
\end{equation}

By using the asymptotically flat solution (\ref{eq:FlatBH}), this equation is a restriction for the radius of stable circular orbits in terms of the parameter $\alpha$,
\begin{equation}
    \omega^2 = \frac{16-18r+6r^2-3\alpha r^3}{6\alpha r^2 (2-r)} >0.
\end{equation}

\begin{figure}[htb!]
	\centering
	\includegraphics[width=1 \linewidth]{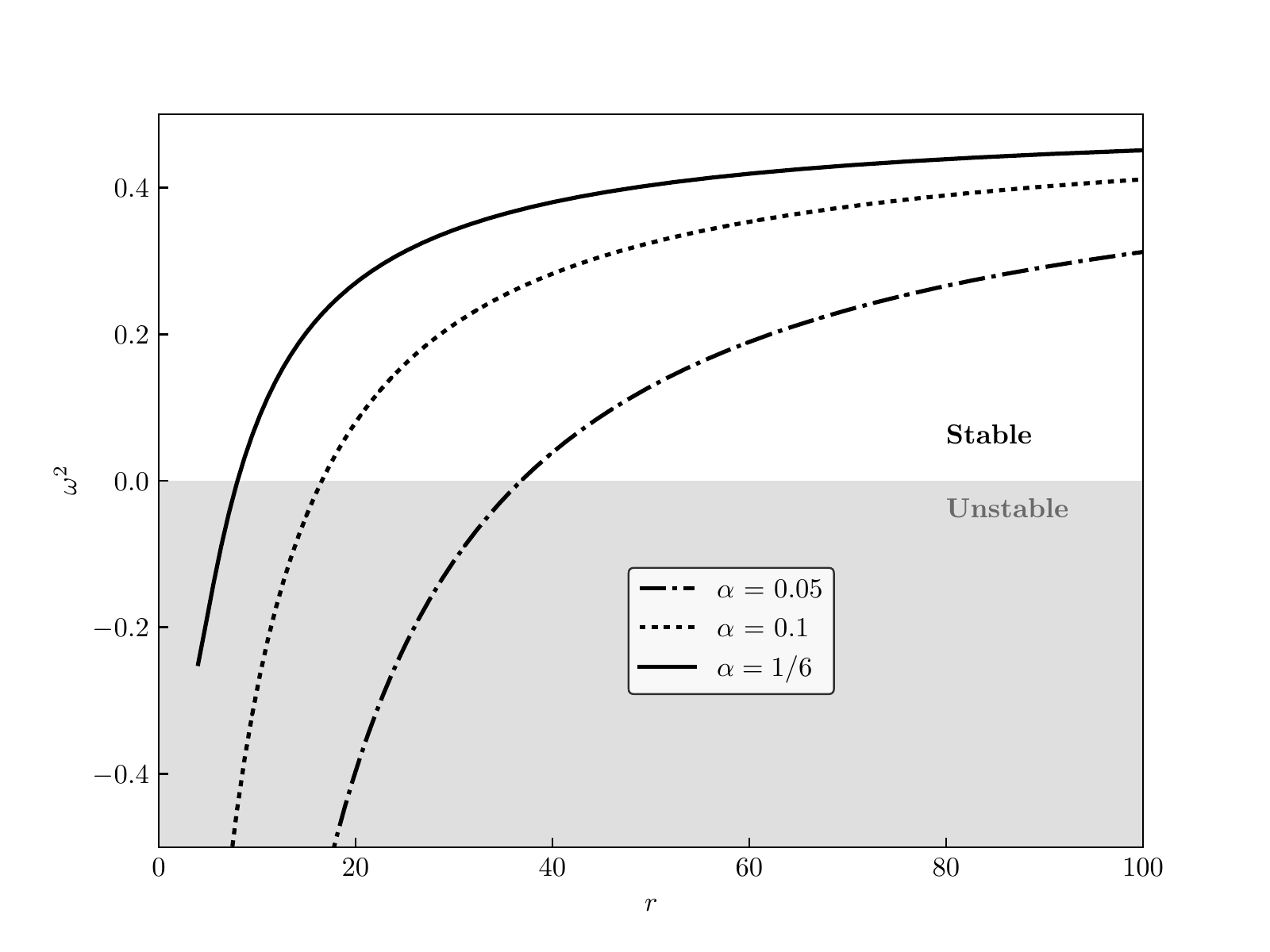}
	\caption{$\omega^{2}$ as function of the radius of the circular orbit for some values of the dimensional parameter $\alpha$ for the asymptotically flat solution. Positive values of $\omega^2$ represent stable circular orbits.} 
	\label{fig:FlatBHStability}
\end{figure}

In Fig. \ref{fig:FlatBHStability} we plot the behavior of $\omega^2$ as function of the radius of the circular orbit for some values of the parameter $\alpha$. Note that the region in which $\omega^2>0$ corresponds to the existence of stable circular orbits.

On the other hand, for asymptotically (anti-)de Sitter black holes described by equation (\ref{eq:AdSBH}) we obtain the condition

\begin{figure}[H]
	\centering
	\includegraphics[width=1 \linewidth]{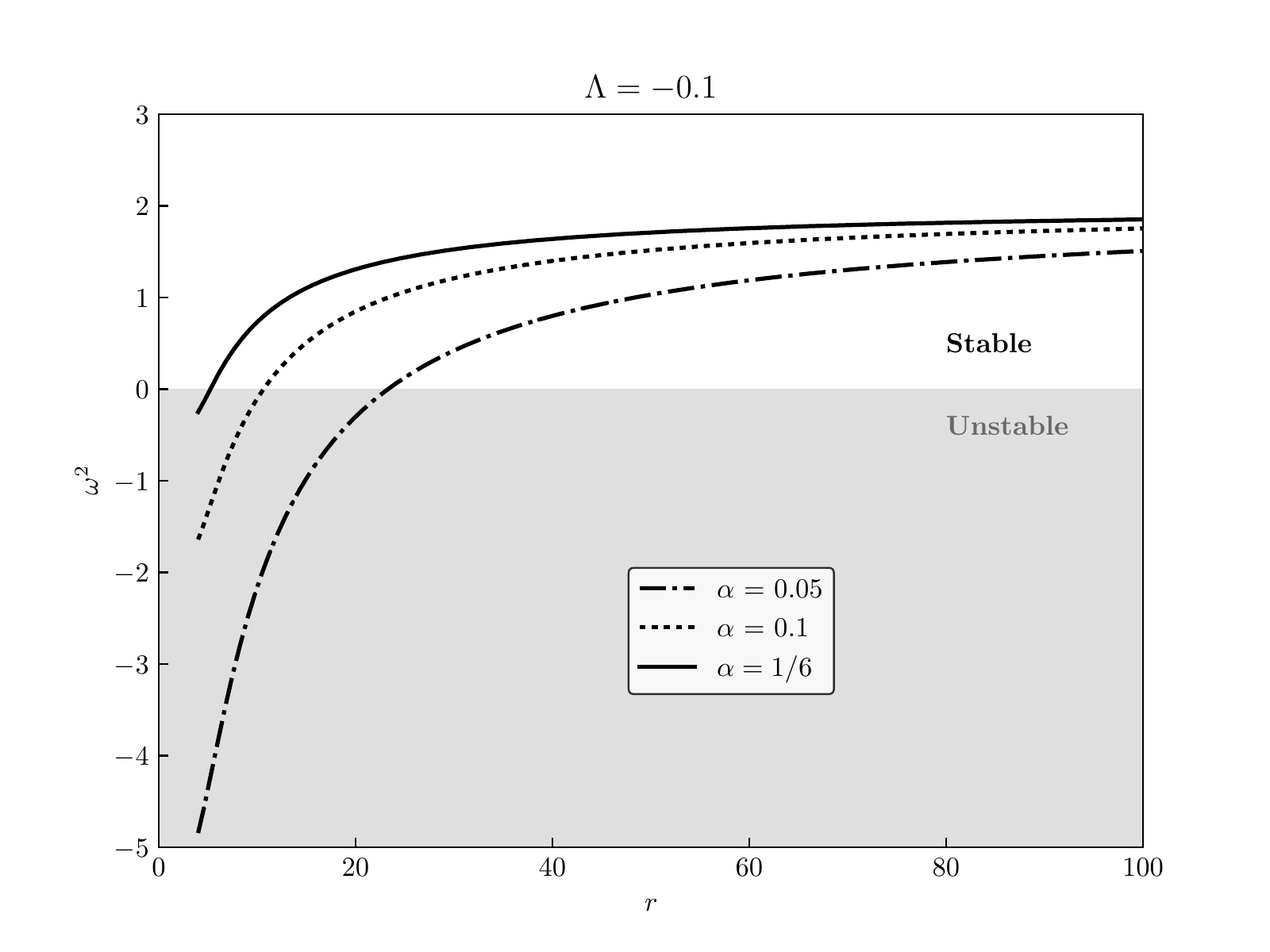}
	\caption{$\omega^{2}$ as function of the radius of the circular orbit for some values of the dimensional parameter $\alpha$ using $\Lambda = - 0.1$. Positive values of $\omega^2$ represent stable circular orbits.} 
	\label{fig:AdSBHStability}
\end{figure}

\begin{widetext}
\begin{equation}
    \omega^2 = \frac{48 \alpha ^2 \Lambda  r^6-60 \alpha  \Lambda  r^5+96 \alpha  \Lambda  r^4-3 \alpha  r^3+6 r^2-18 r+16}{6 \alpha  r^2 \left(4 \alpha  \Lambda  r^4-r+2\right)} > 0
\end{equation}
\end{widetext}
which is plotted in Fig. \ref{fig:AdSBHStability} for the particular value $\Lambda = -0.1$ and for some values of the parameter $\alpha$.\\

It is important to note that it is possible to obtain the radius of the ISCO by obtaining numerically the zero of the function $\omega^2$ for each value of $\alpha$. The result coincides with the ISCO radius shown in Figure \ref{fig:BHrISCO}.

\subsection{Geodesic motion in the equatorial plane}
To study the geodesic motion of a test particle in the equatorial plane, we obtain the equation of the orbit by introducing a new coordinate $u=1/r(\phi)$ and rewriting (\ref{eq:radialequation}) in terms of the azimuthal angle $\phi$ and the constant of motion $l=r^{2}\dot{\phi}$. This gives the first order differential equation
\begin{equation}
  \left( \dfrac{du}{d\phi} \right)^{2} = \dfrac{1}{l} \left[ E^{2} - B(u)(1+u^{2}l^{2}) \right]  \ 
  \label{eq:firstOrder}
\end{equation}
which, after differentiating with respect to $\phi$ and dividing by $2 du/d\phi$, gives the second order  differential equation of motion 
\begin{equation}
\dfrac{d^{2}u}{d\phi^{2}} = - \left[ u B(u) + \dfrac{1}{2}\left( \dfrac{1}{l^{2}} + u^{2} \right)\dfrac{dB}{du} \right] \ .
\label{eq:orbitEq}
\end{equation}

Using (\ref{eq:AdSBH}) for $B(u)$, we get 
\begin{align}
\nonumber
\dfrac{d^{2}u}{d\phi^{2}} &= -\left( \dfrac{1}{2} + \dfrac{1}{3\alpha l^{2}}\right)u(\phi) \, + \, \dfrac{1}{2\alpha}u^{2}(\phi)\\[2ex]
    & \ \ \ \, - \,  \dfrac{2}{3\alpha}u^{3}(\phi) \, - \, \dfrac{2 \Lambda}{3l^{2}u^{3}(\phi) } \, + \, \dfrac{1}{6\alpha l^{2}} \ .
    \label{eq:Orbit}
\end{align}

This equation looks like the orbital equation obtained in general relativity for Schwarzschild's or Reissner-Nördstrom solutions in the presence of a cosmological constant. However, it is important to remember that it is not possible to obtain a general relativity limit of this relation because the parameter $\alpha$ can not be zero. 

\begin{figure}[htb!]
	\centering
	\includegraphics[width=1 \linewidth]{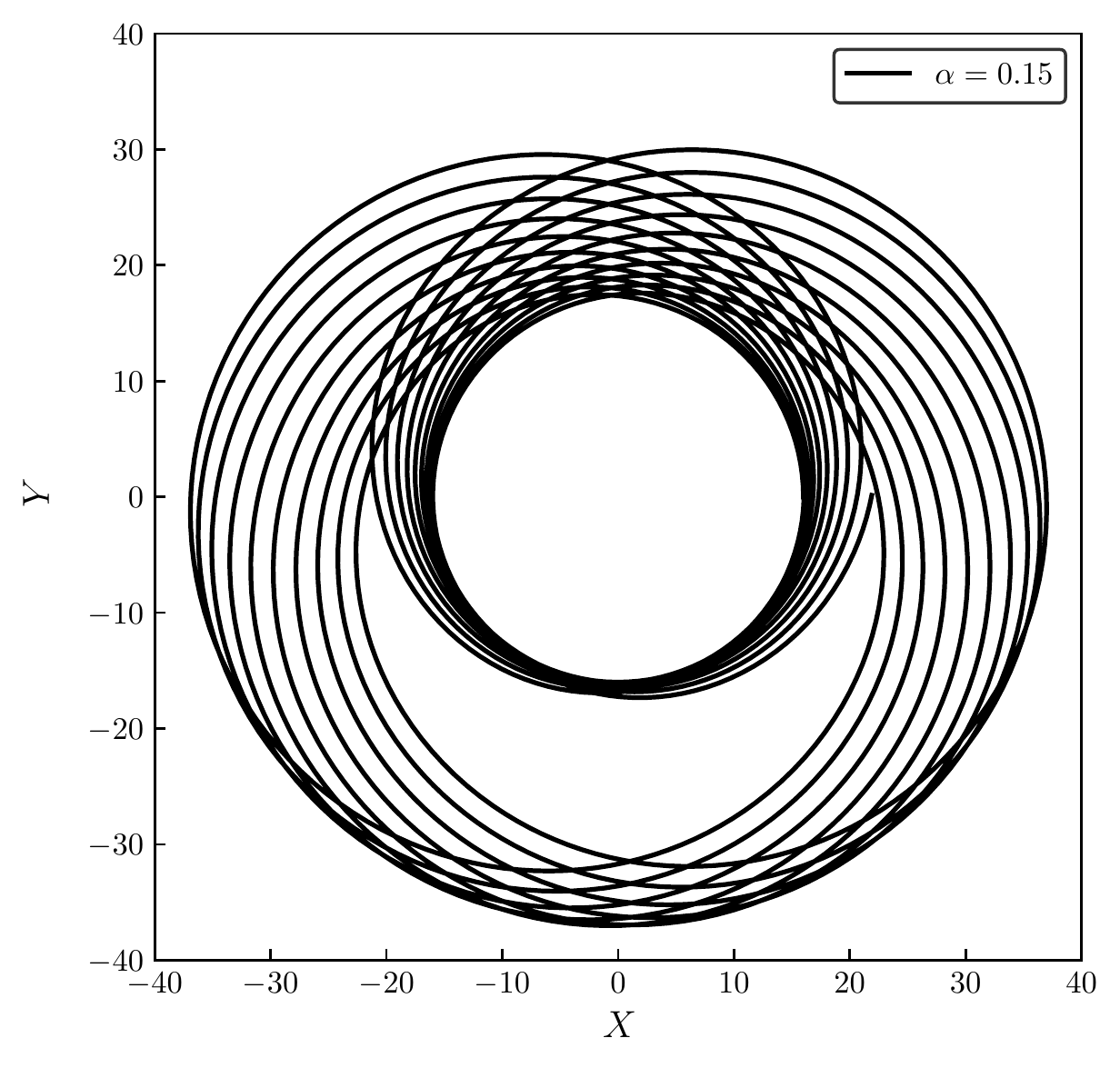}
	\caption{Geodesic motion of a test particle moving in the equatorial plane of the asymptotically flat solution background when $\alpha=0.15$ after 20 turns. The initial conditions for this solution are $u(0)=1/16$ and $u'(0)=0$ with a specific angular momentum $l=8$.} 
	\label{fig:flatGeodesic}
\end{figure}

\begin{figure}[h!]
	\centering
	\includegraphics[width=1 \linewidth]{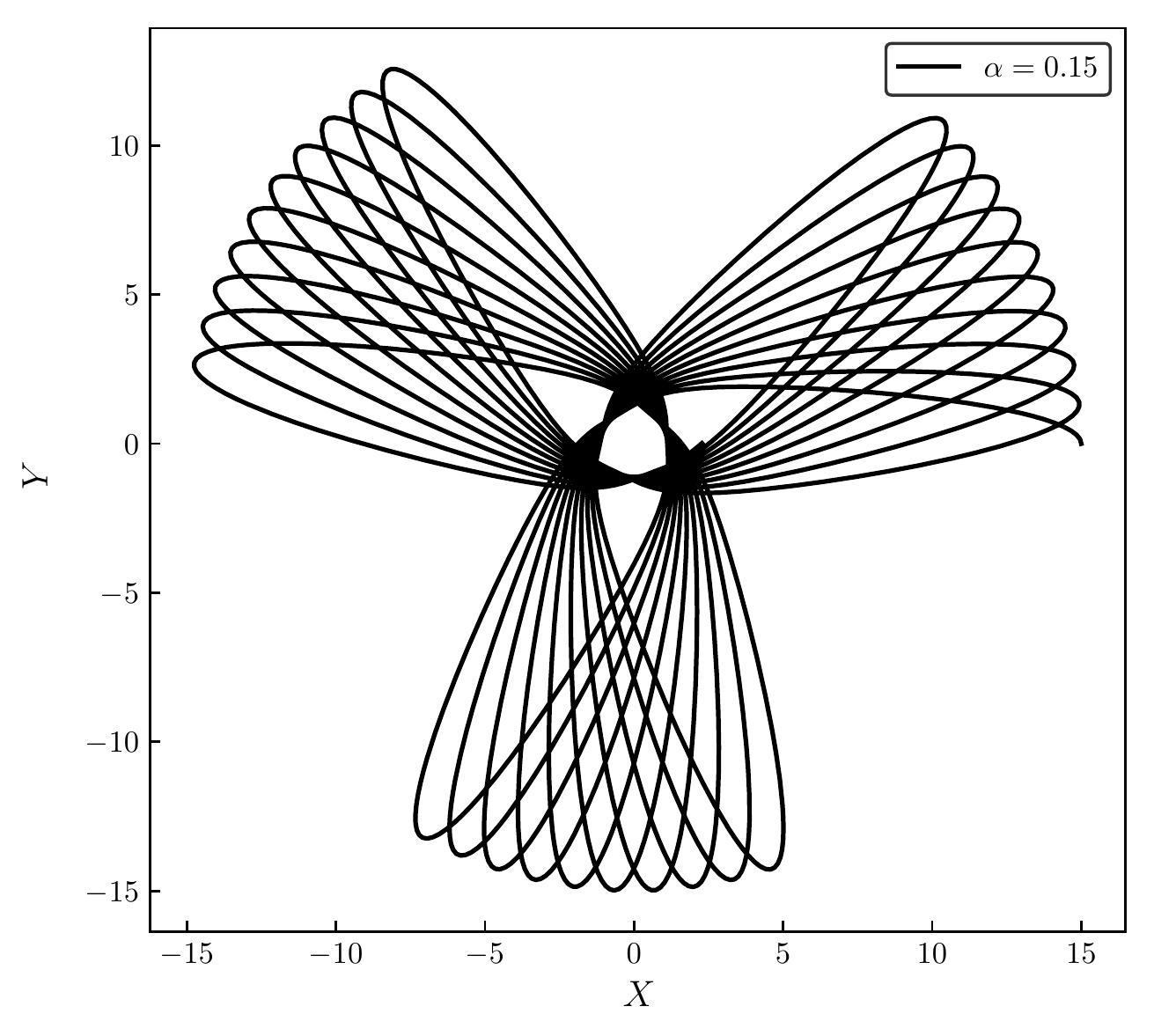}
	\caption{Geodesic motion of a test particle moving in the equatorial plane of the asymptotically (anti-)de Sitter solution background when $\alpha=0.15$ and $\Lambda=-0.1$ after 20 turns. The initial conditions for this solution are $u(0)=1/15$ and $u'(0)=0$ with a specific angular momentum $l=8$.} 
	\label{fig:AdSGeodesic}
\end{figure}

Numerical solutions of (\ref{eq:Orbit}) for the asymptotically flat and (anti-)de Sitter metrics when $\alpha = 0.15$ give the bound orbits shown in Figures \ref{fig:flatGeodesic} and \ref{fig:AdSGeodesic}.

\section{The Stationary Axisymmetric Charged Black Holes}
In this section we'll apply the Janis-Newman algorithm modification proposed by Azreg-Ainou in \cite{Azreg2014}, to obtain a stationary spacetime from a static metric without the complexification step. We start by writing the metric (\ref{eq:AdSBH}) in advanced null coordinates $(u,r,\theta,\phi)$ using the transformation $du = dt - \frac{dr}{B(r)}$. This gives
\begin{equation}
    ds^2 = B(r) du^2 + 2dudr - r^2 (d\theta^2 + \sin^2 \theta d\phi^2). \label{eq:EFstaticmetric}
\end{equation}

Using the tetrad vectors
\begin{equation}
    l^\mu = \delta^\mu_r\, , \, n^\mu = \delta^\mu_u - \frac{B(r)}{2} \delta^\mu_r \, , \, m^\mu = \frac{1}{\sqrt{2}r}\left( \delta^\mu_\theta + \frac{i}{\sin \theta} \delta^\mu _\phi \right)
\end{equation}
it is possible to write the metric (\ref{eq:EFstaticmetric}) as
\begin{equation}
    g^{\mu \nu} = l^\mu n^\nu + l^\nu n^\mu - m^\mu \bar{m}^\nu - m^\nu \bar{m}^\mu
\end{equation}
where $\bar{m}^\mu$ is the complex conjugate of $m^\mu$. The next step is to introduce a complex transformation in the $u-r$ plane,
\begin{align}
    r = &r' - i a \cos \theta  \\
    u = &u' + i a \cos \theta ,
\end{align}
where $a$ will be interpreted as the spin parameter of the rotating solution. In the original Janis-Newman algorithm, it is necessary to the complexify  the radial coordinate. However, the Azreg-Ainou method drops this step by introducing three functions, $C_1=C_1(r,a,\theta)$, $C_2=C_2(r,a,\theta)$ and $H=H(r,a,\theta)$, such that the new metric will be written in terms of a new set of null tetrads given by
\begin{align}
    l'^\mu = &\delta^\mu_r \\
    n'^\mu = &\sqrt{\frac{C_2}{C_1}} \delta^\mu_u - \frac{C_2}{2} \delta^\mu_r \\
    m^\mu = &\frac{1}{\sqrt{2H}}\left[ ia\sin \theta (\delta^\mu_u - \delta^\mu_r)+ \delta^\mu_\theta + \frac{i}{\sin \theta} \delta^\mu _\phi \right].
\end{align}

The line element that is recovered from this tetrad is
\begin{align}
    ds^2 = & C_1 du^2 - 2 \sqrt{\frac{C_1}{C_2}} dudr - 2a \sqrt{\frac{C_1}{C_2}} \sin^2 \theta drd\phi \notag \\
    &+ 2a\sin^2 \theta \left(C_1 - \sqrt{\frac{C_1}{C_2}}\right) du d\phi + H d\theta^2 \notag \\
    & \sin^2 \theta \left[ H + \left(2\sqrt{\frac{C_1}{C_2}} - C_1 \right) a^2 \sin ^2 \theta \right]d\phi^2 .
\end{align}

In order to write this metric in Boyer-Linquist coordinates, we introduce the transformation
\begin{align}
    du = &dt' -\frac{r^2 + a^2}{B(r) r^2 + a^2} dr \\
    d\phi = &d\phi ' - \frac{a}{B(r) r^2 + a^2} dr.
\end{align}

Finally, the unknown functions introduced above will be determined by choosing $H= r^2 + a^2 \cos^2 \theta$ and by imposing that the cross-term $dtdr$ in the metric vanishes. We obtain
\begin{align}
    C_1 (r,a,\theta) = C_2(r,a,\theta)= \frac{B(r) r^2 + a^2 \cos^2 \theta}{r^2 + a^2 \cos^2 \theta}
\end{align}
and therefore, the rotating metric is
\begin{align}
    ds^2 = & \frac{\Delta - a^2 \sin^2 \theta}{\Sigma} dt^2 + 2\left( \frac{r^2 + a^2 - \Delta}{\Sigma} \right) a \sin^2 \theta dtd\phi \notag \\
    & - \frac{\Sigma}{\Delta} dr^2 - \Sigma d\theta^2 \notag \\
    &- \left( \frac{(r^2 + a^2)^2 - \Delta a^2 \sin^2 \theta}{\Sigma}\right) \sin^2\theta d\phi^2,\label{eq:RotatingBH}
\end{align}
where
\begin{equation}
    \Sigma = r^2 + a^2 \cos^2 \theta
\end{equation}
and
\begin{equation}
    \Delta = \frac{1}{2}\Sigma - \frac{2}{3} r^4 \Lambda - \frac{r}{3\alpha} + \frac{1}{3\alpha} + a^2 \sin^2 \theta.
\end{equation}

Using the metric for this stationary solution, we obtain the Ricci curvature invariant
\begin{equation}
    R = \frac{1+8\Lambda r^2 -\frac{18 a^4 \alpha^2 \cos^2\theta \sin^2 \theta }{\left(3 a^2 \alpha \left(\cos^2\theta-2\right)+r \left(4 \alpha \Lambda  r^3-3 \alpha r+2\right)-2\right)^2}}{r^2 + a^2 \cos^2 \theta},
\end{equation}
 which correctly recovers the Ricci invariant for the static black hole, given in equation (\ref{eq:RicciScalar}), when $a=0$. \\


The spacetime in equation (\ref{eq:RotatingBH}) is singular at the surface defined by $\Delta = 0$. In general, this is a fourth order algebraic equation which has two real roots (inner and outer horizons) and two complex roots. For a spacetime with vanishing cosmological constant, this relation reduces to a quadratic equation, an it is possible to solve analytically for the radii of the horizons,
\begin{equation}
    r_{\pm} = \frac{1}{3\alpha}\left[ 1\pm \sqrt{1- 6\alpha -\frac{9}{2} a^2 \alpha^2(3 - \cos 2\theta)} \right]. \label{eq:rotatingHorizon1}
\end{equation}

The most important feature of this result is that the event horizon of this black hole, $r_H = r_+$, is a function depending on the angle $\theta$. This will imply, as discussed below, that the Hawking temperature is also angle-dependent. Although this is not the typical and/or expected behavior, the possibility of angle dependence in these and other physical properties have been studied before. For example, angle dependence on the Hawking temperature of general non-stationary black holes is studied in \cite{Yang1995} while the behavior of particular cases such as the non-stationary Kerr black hole is reported in \cite{Wu1993} or for the non-stationary Kinnersley black hole in \cite{Luo1993}. \\

\begin{figure}[htb!]
	\centering
	\includegraphics[width=1 \linewidth]{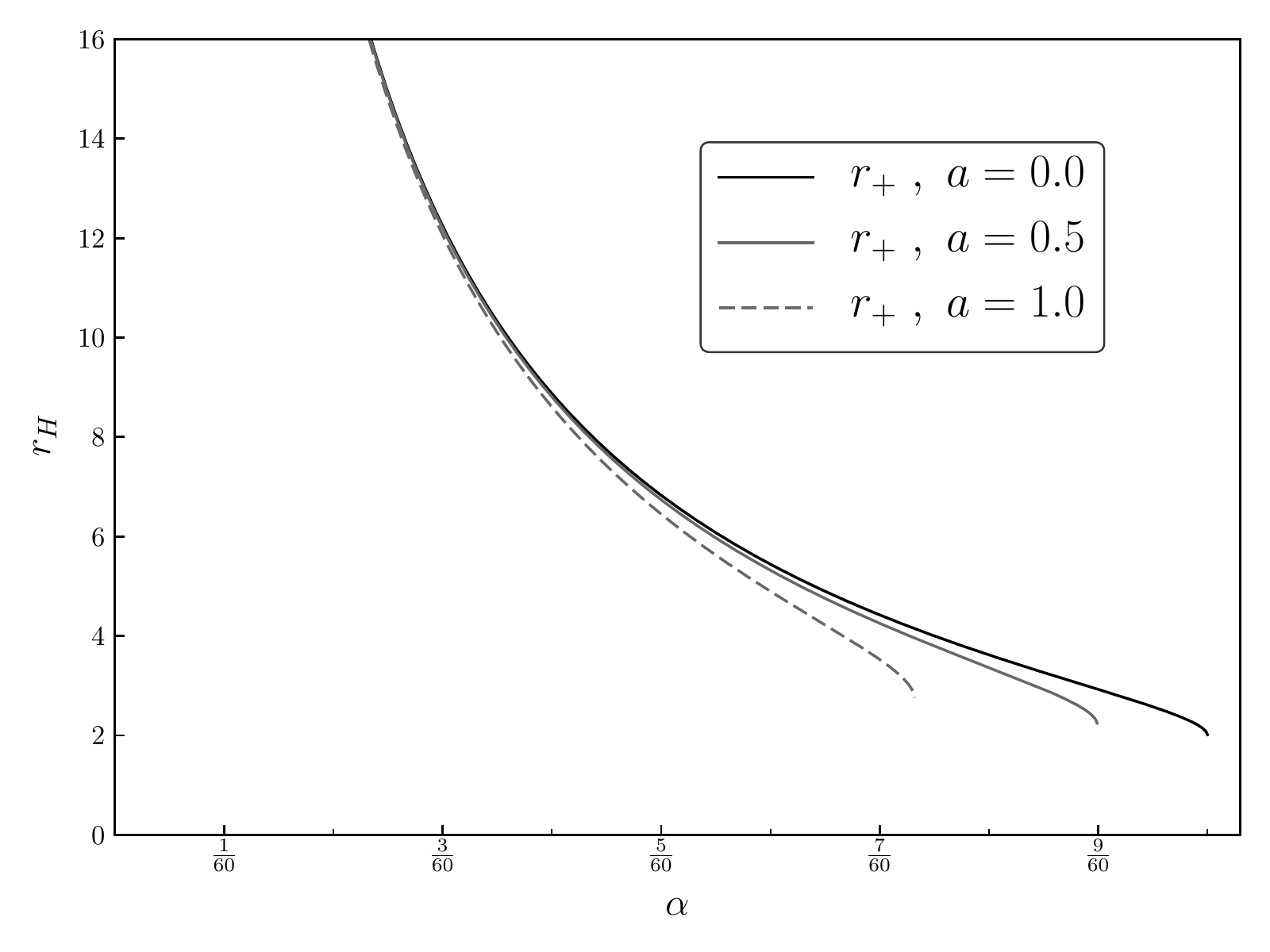}
	\caption{Radius of the event horizon as function of the parameter $\alpha$ for the asymptotically flat rotating black hole when $\theta=\pi/2$.} 
	\label{fig:HorizonRot}
\end{figure}

The behavior of the radius of the event horizon depending on the parameters $\alpha$ and $a$ for the inclination angle $\theta=\frac{\pi}{2}$, is shown in Figure \ref{fig:HorizonRot}. We have shown above that the horizons for the static solution, $a=0$, become a degenerate horizon when $\alpha=\frac{1}{6}$. Equation (\ref{eq:rotatingHorizon1}) generalizes this result so that, for a given value of the spin parameter $0<a\leq 1$, the extremal black hole is obtained when the parameter $\alpha$ takes it maximum value
\begin{equation}
    \alpha_{max} = \frac{2-\sqrt{2} \sqrt{2-a^2 (\cos 2 \theta -3)}}{3 a^2 \left(\cos 2 \theta -3\right)}.
\end{equation}

The radius of the degenerate horizon is 
\begin{equation}
    r^{ext}_H = \frac{a^2 ( \cos 2 \theta -3 )}{2-\sqrt{2} \sqrt{2-a^2 (\cos 2 \theta -3)}}.
\end{equation}

\begin{figure}[htb!]
	\centering
	\includegraphics[width=1 \linewidth]{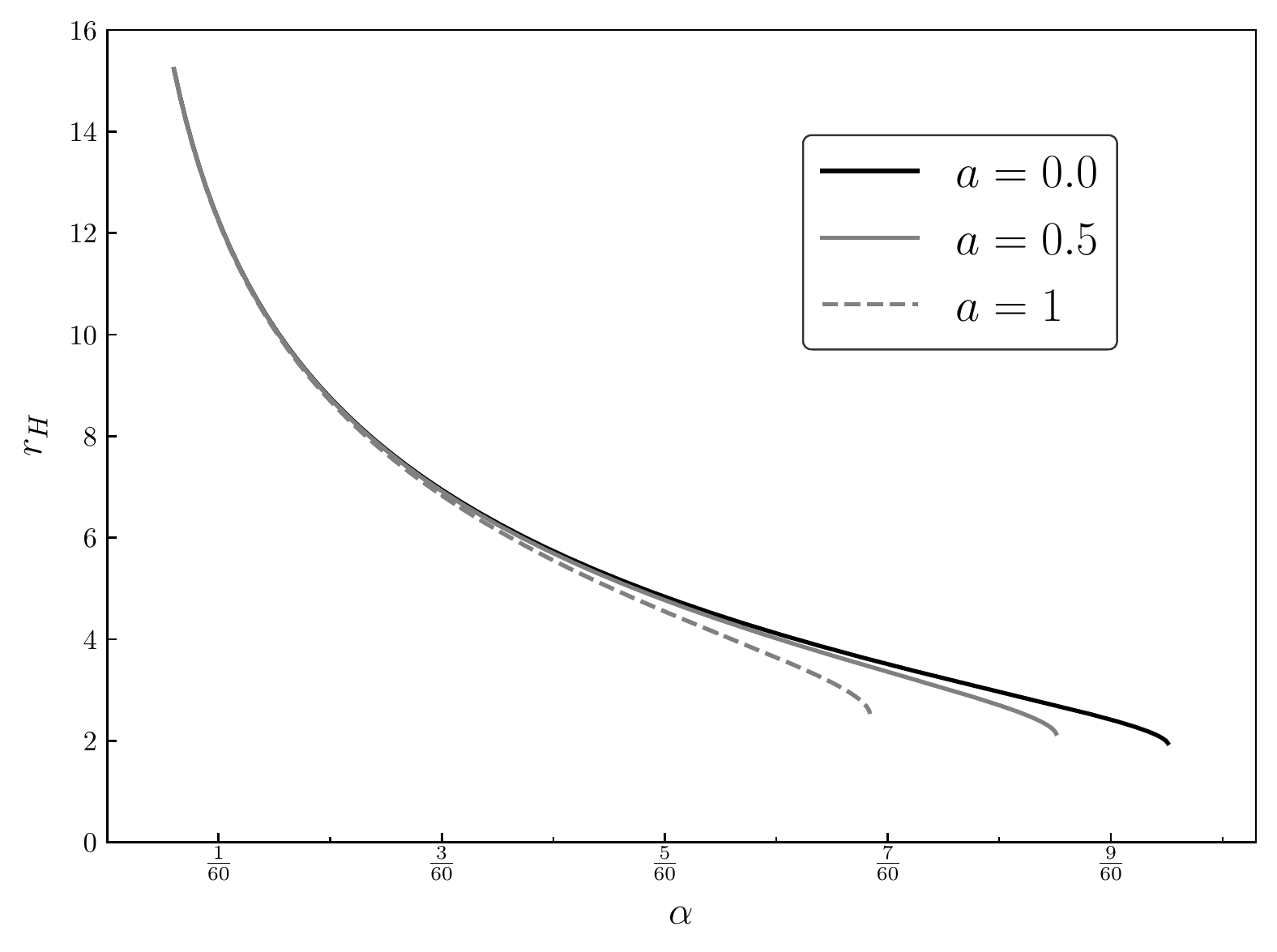}
	\caption{Radius of the event horizon as function of the parameter $\alpha$ for the asymptotically (anti-)de Sitter rotating black hole for $\Lambda=-0.01$ and $\theta=\pi/2$.} 
	\label{fig:HorizonRotLambda}
\end{figure}

\begin{figure}[htb!]
	\centering
	\includegraphics[width=1 \linewidth]{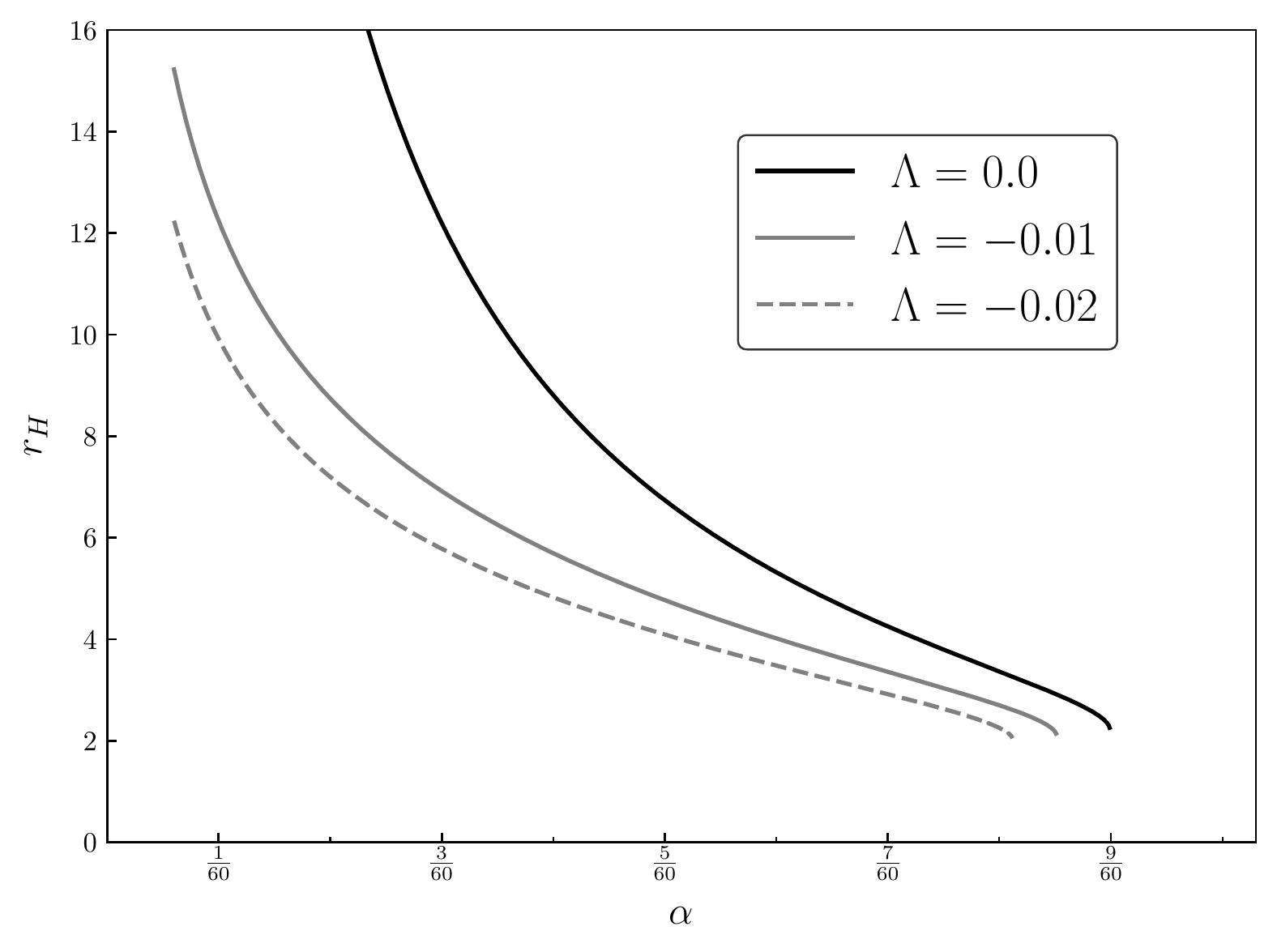}
	\caption{Radius of the event horizon as function of the parameter $\alpha$ for the asymptotically (anti-)de Sitter rotating black hole for $a=0.5$ and $\theta=\pi/2$.} 
	\label{fig:HorizonRotLambdaB}
\end{figure}

In the case of a black hole with non-vanishing cosmological constant, it is also possible to solve analytically the equation $\Delta=0$ to obtain the radius of both inner and outer horizon. However, the expressions are cumbersome and not particularly illuminating. Therefore, we prefer to show the plots of the horizon radius depending on $a$ and $\alpha$ in Figures \ref{fig:HorizonRotLambda} and \ref{fig:HorizonRotLambdaB}, to note that the behavior is completely similar to our previous results.

\subsection{Hawking Temperature of the Rotating Black Hole}

The spacetime represented by the metric (\ref{eq:RotatingBH}) admits the Killing vectors $\xi^\mu = \partial_t$ and $\zeta^\mu = \partial_\phi$, associated with time translation and rotational invariance, respectively. Defining the Killing vector $\chi^\mu = \xi^\mu + \Omega \zeta^\mu$, we obtain the angular velocity at the horizon, $\Omega_+$, by imposing $\chi^\mu$ to be a null vector at the surface $\Delta=0$. This gives the value
\begin{equation}
    \Omega_+ =\frac{a}{r_+^2 + a^2}. 
\end{equation}

The Hawking temperature associated with this black hole can be easily obtained through the relation
\begin{equation}
    T = \frac{\kappa}{2\pi},
\end{equation}
where the surface gravity is
\begin{equation}
    \kappa^2 = - \frac{1}{2} \nabla^\mu \chi ^\nu  \nabla_\mu \chi _\nu.
\end{equation}

\begin{figure}[htb!]
	\centering
	\includegraphics[width=1 \linewidth]{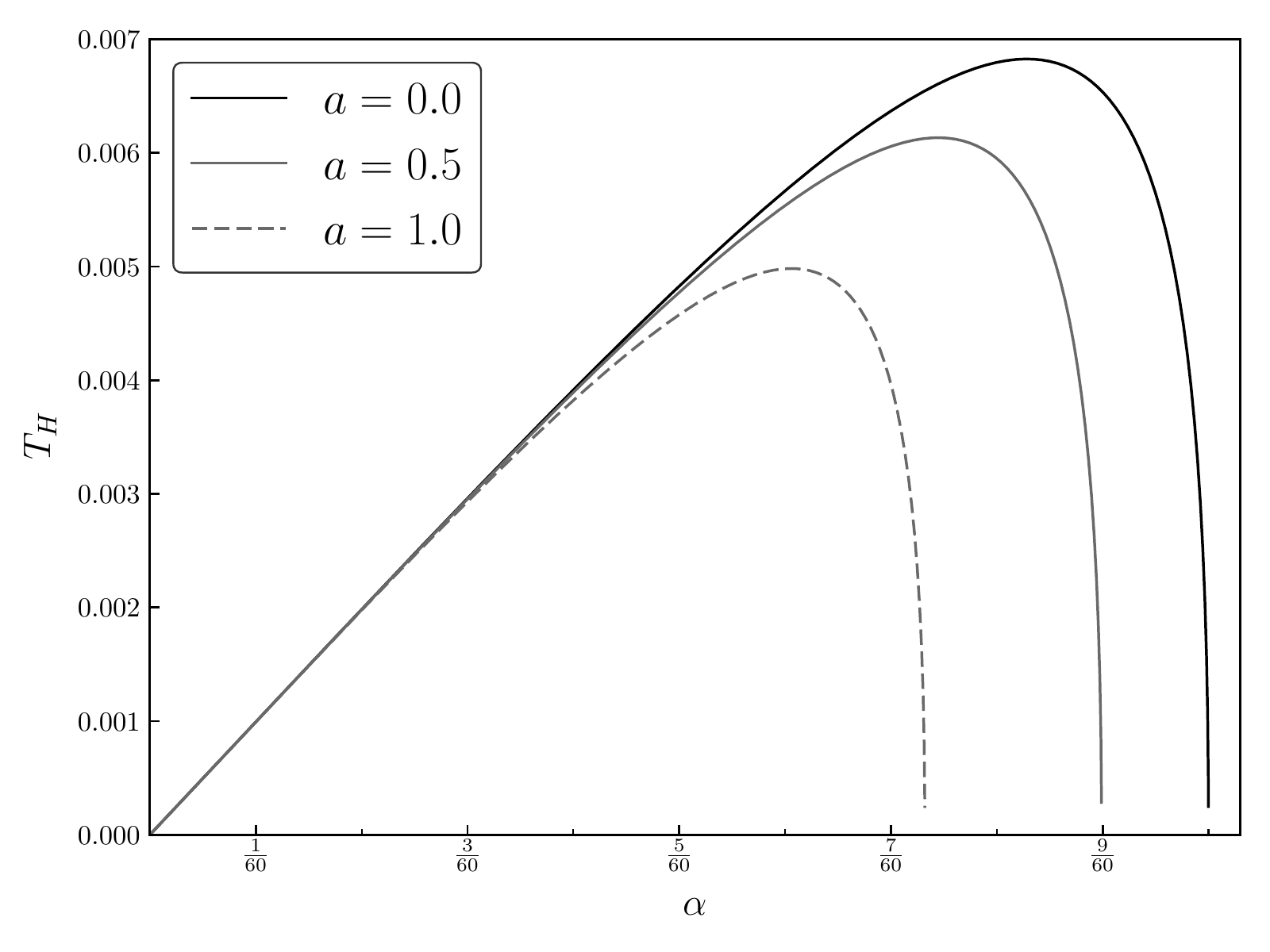}
	\caption{Hawking temperature as function of $\alpha$ for the asymptotically flat rotating black hole for $\theta=\pi/2$.} 
	\label{fig:TempRot}
\end{figure}

\begin{figure}[htb!]
	\centering
	\includegraphics[width=1 \linewidth]{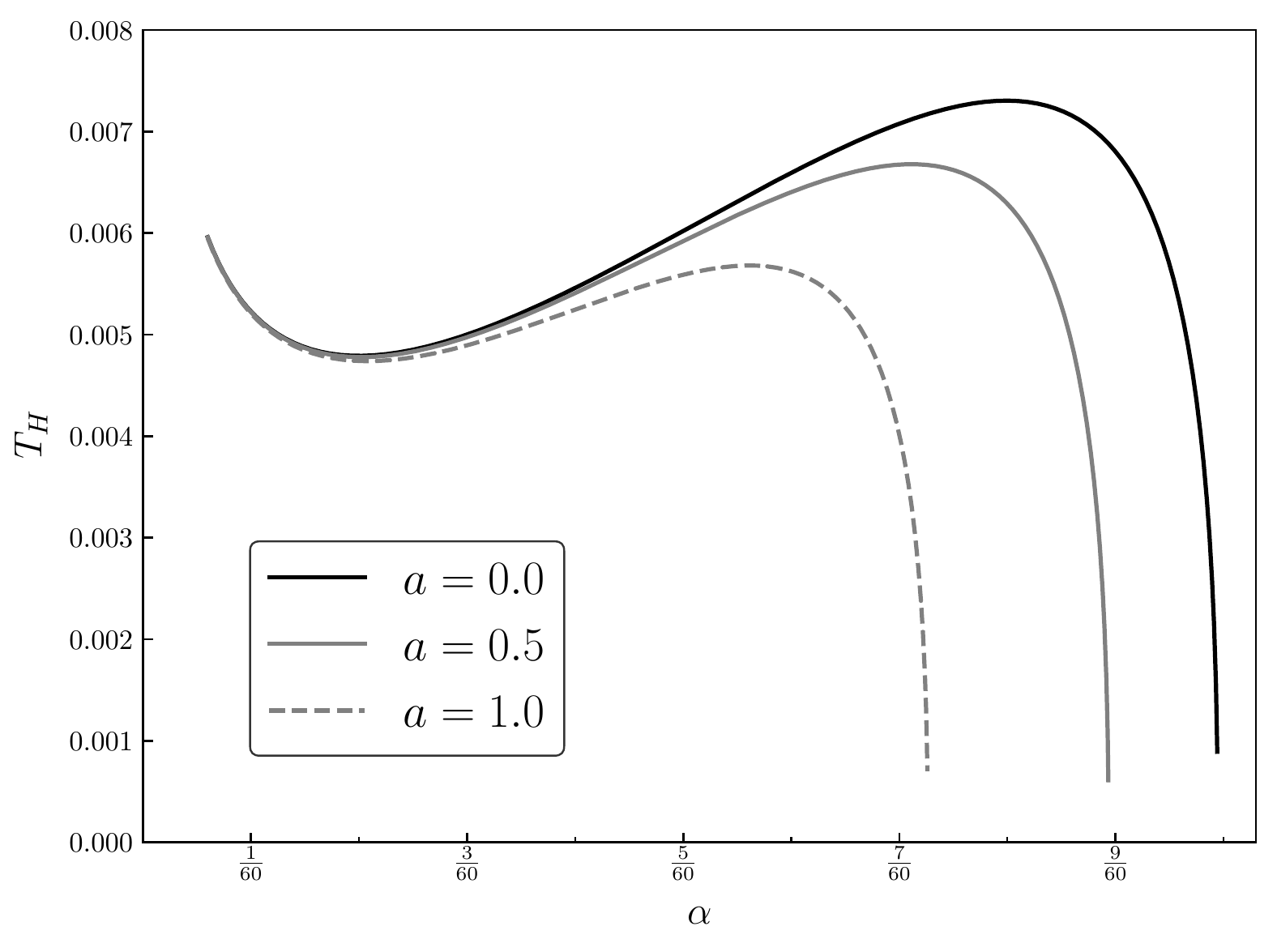}
	\caption{Hawking temperature as function of $\alpha$ for the asymptotically (anti-)de Sitter rotating black hole for $\Lambda=-0.001$ and $\theta=\pi/2$.} 
	\label{fig:TempLambda1}
\end{figure}

This gives the value of the temperature at the event horizon of
\begin{equation}
    T_+ = \frac{1}{4\pi(r_+^2 + a^2)} \left. \frac{d \Delta}{dr} \right|_{r=r_+} = \frac{r_+-\frac{8}{3}r_+^3\Lambda - \frac{1}{3\alpha} }{4\pi (r_+^2 + a^2)},
\end{equation}
which clearly depends on the angle $\theta$ through the value of $r_+$. Hence, we can expect an angle-dependent thermal radiation spectrum, similar to those reported for other black hole solutions in \cite{Yang1995}. Figures \ref{fig:TempRot} and \ref{fig:TempLambda1} show that the behavior of the temperature at a fixed angle $\theta$  is very similar to the temperature of the non rotating solutions described above.

\begin{figure}[htb!]
	\centering
	\includegraphics[width=1 \linewidth]{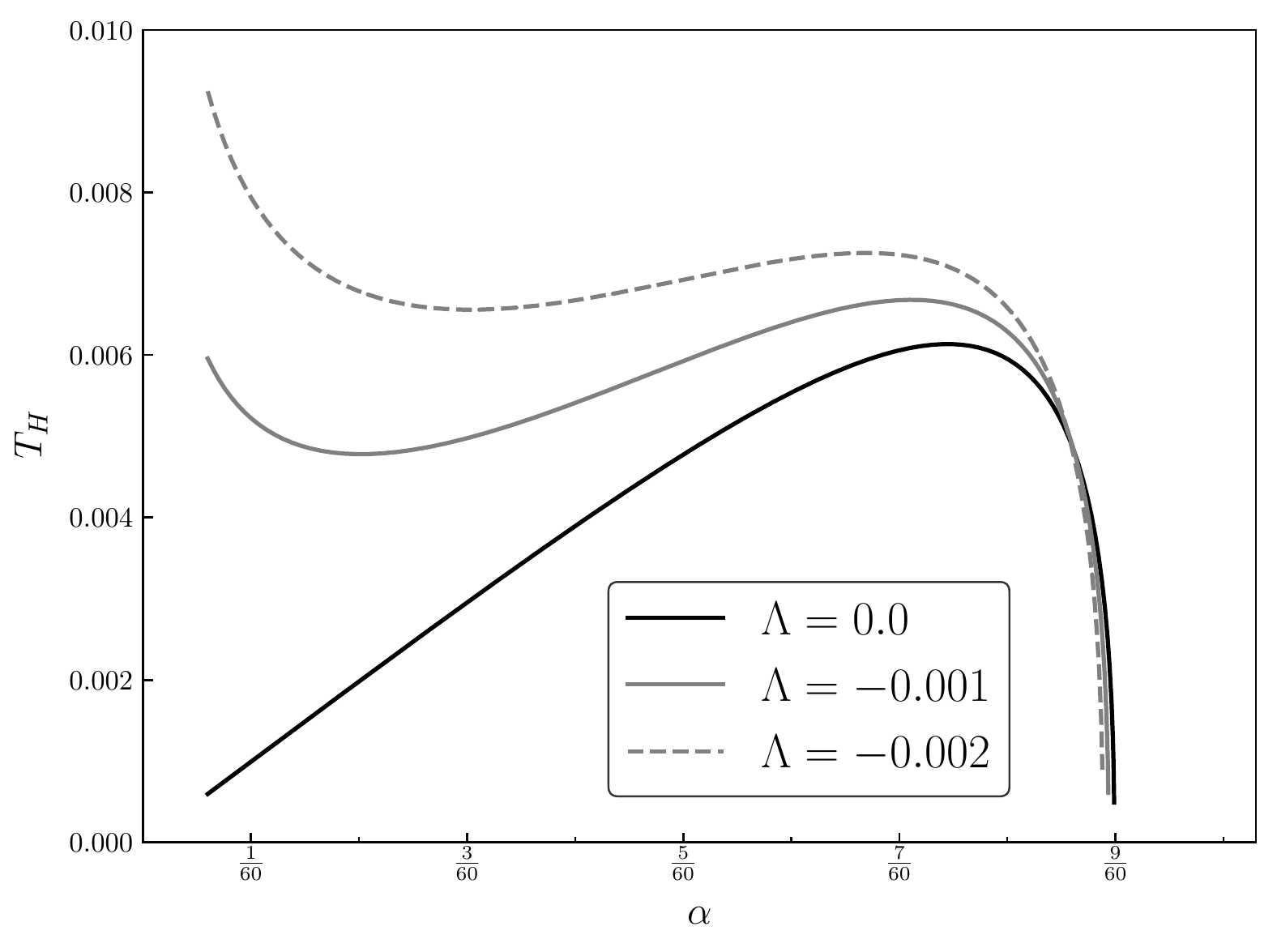}
	\caption{Hawking temperature as function of $\alpha$ for the asymptotically (anti-)de Sitter rotating black hole for $a=0.5$ and $\theta=\pi/2$.} 
	\label{fig:TempLambdaX}
\end{figure}

\subsection{Ergosphere}

The stationary limit surface or ergosphere, is defined by the equation $g_{tt} (r_{st}, \theta) = 0$, which gives, in this case,
\begin{equation}
    \frac{1}{2}r_{e}^2 + \frac{1}{2}a^2 \cos^2 \theta - \frac{2}{3} r_{e}^4 \Lambda - \frac{r_{e}}{3\alpha} + \frac{1}{3\alpha} = 0. 
\end{equation}

A plot of $r_e$ shows that the ergosphere and the event horizon meet at the poles but not at the equatorial plane. Hence, there is a region between the horizon and the stationary limit surface, called the ergoregion.

\begin{figure}[htb!]
	\centering
	\includegraphics[width=1 \linewidth]{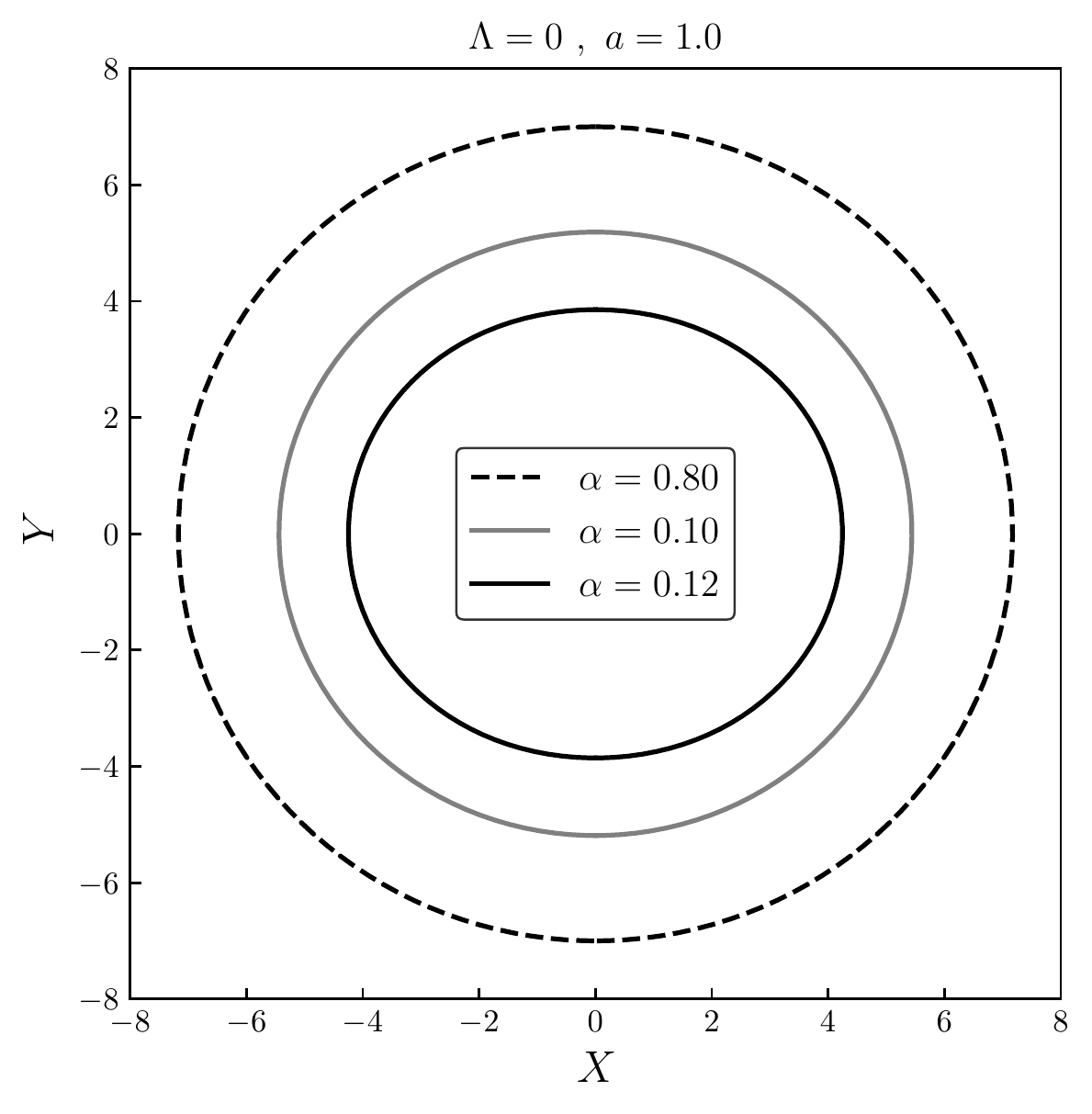}
	\caption{Ergosphere for the asymptotically flat rotating black hole for the rotating parameter $a=1.0$.} 
	\label{fig:ErgoEsferaFlat}
\end{figure}

\begin{figure}[htb!]
	\centering
	\includegraphics[width=1 \linewidth]{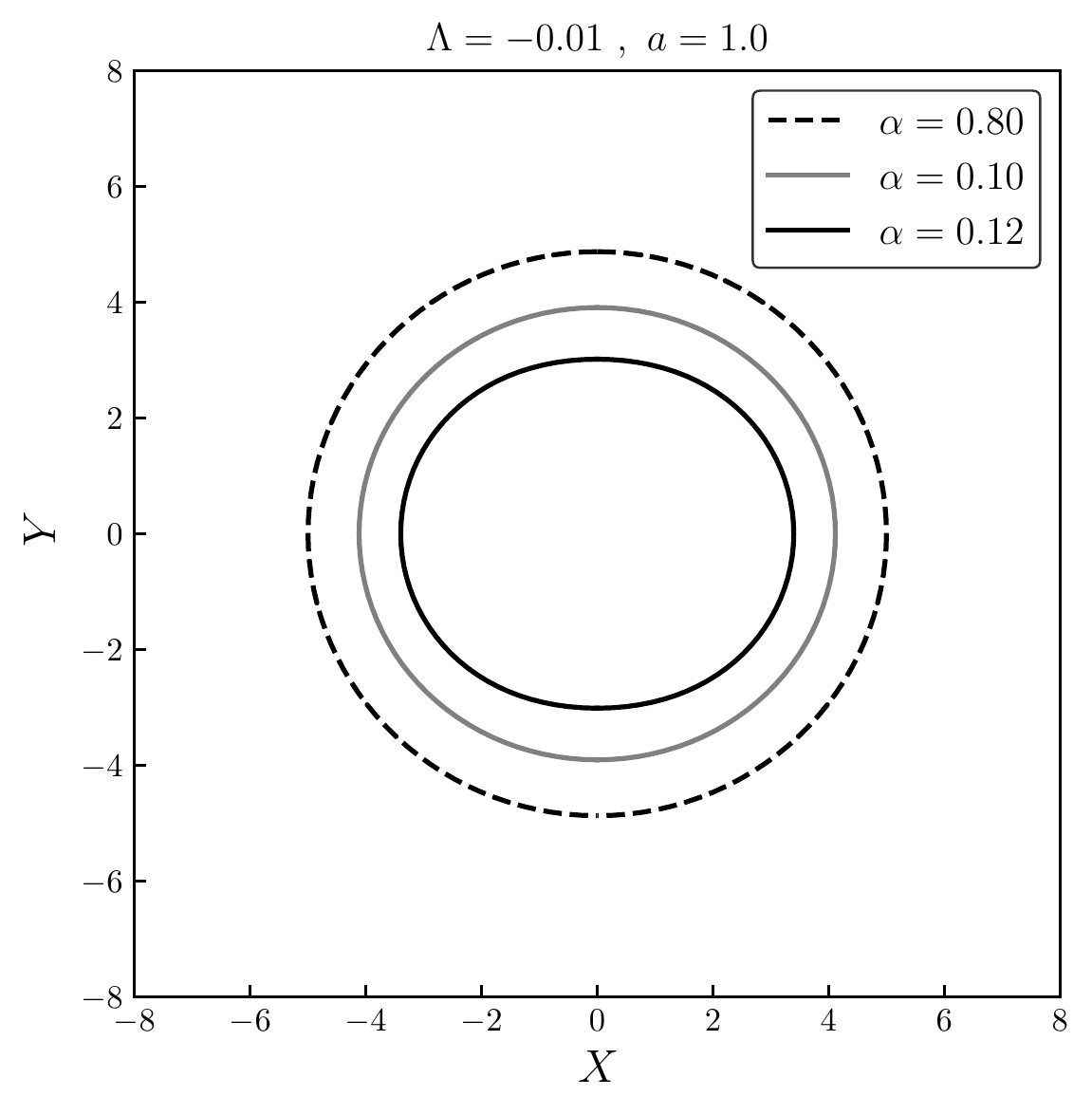}
	\caption{Ergosphere for the asymptotically (anti-)de Sitter rotating black hole for the rotating parameter $a=1.0$ and $\lambda=-0.01$.} 
	\label{fig:ErgoEsferaFlat}
\end{figure}

\begin{figure}[htb!]
	\centering
	\includegraphics[width=1 \linewidth]{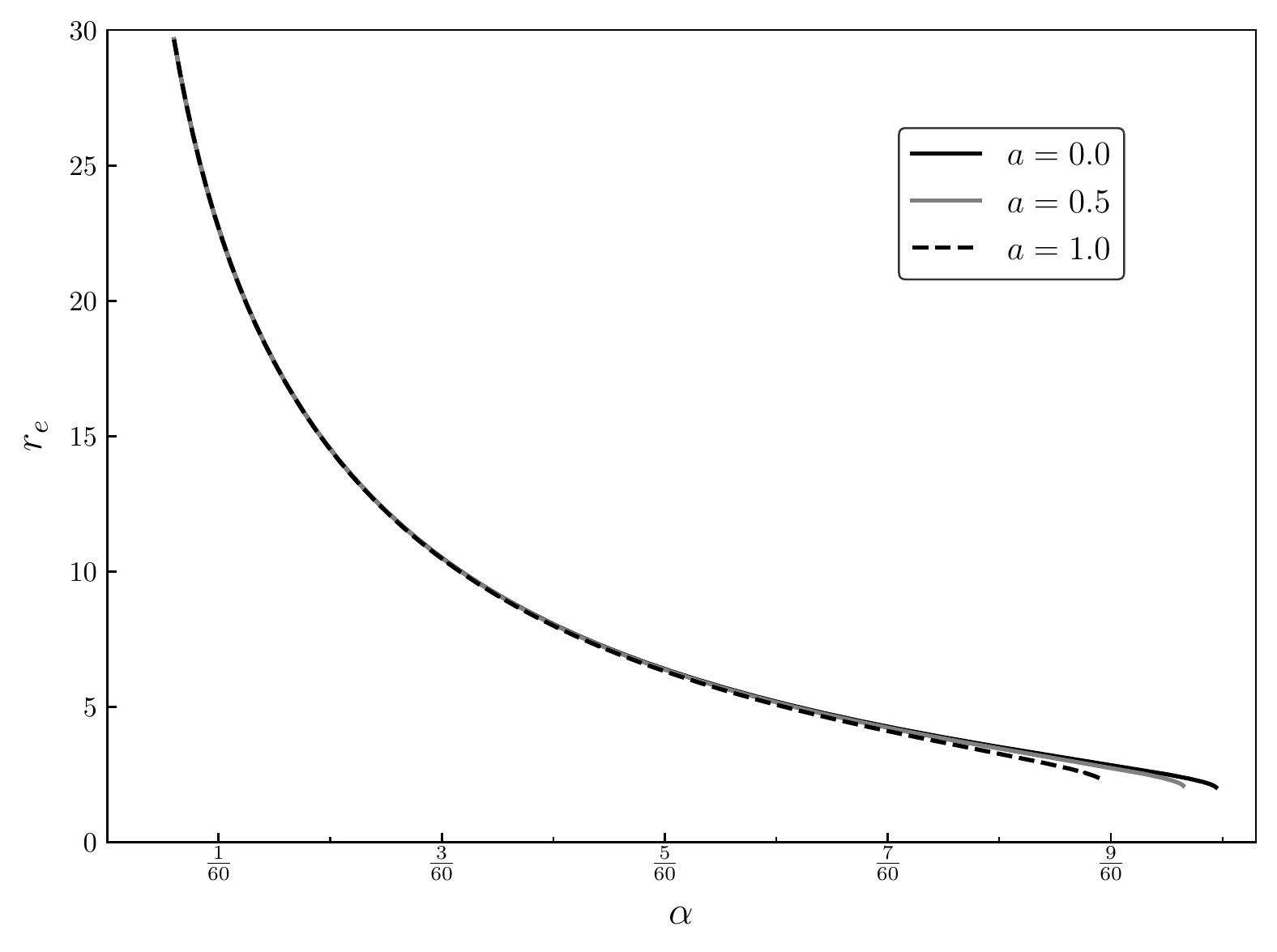}
	\caption{Ergosphere radius for $\theta=\pi/4$ for the asymptotically (anti-)de Sitter rotating black hole for the rotating parameter $a=1.0$ and $\lambda=-0.001$.} 
	\label{fig:ErgoEsferaFlat}
\end{figure}

\section{Conclusion}

The charged, spherically symmetric solutions depending on a dimensional parameter $0<\alpha < \frac{1}{6}$ reported by Nashed and Capozziello in \cite{Nashed2019} satisfy the Maxwell-$f(R)$ field equations in the presence of a cosmological constant. In this paper we have obtained the correct description of the physical and thermodynamical properties associated with these black hole type solutions, including radius of the horizon, Hawking temperature, entropy, quasi-local energy and Gibbs free energy. The behavior of these quantities show that the black holes evaporate similarly to the Reissner-Nördstrom process. In particular, we show that the evaporation process is well-behaved and  terminates in a frozen remnant with a final horizon radius $r_H^{ext} = 2$. Accordingly, the entropy and the Gibbs free energy both have a smooth behavior, showing no phase transitions for the black holes in the allowed range of the parameter $\alpha$. This result is significantly different from that obtained in \cite{Nashed2019}, where the authors reported a non-existing phase transition due to a wrong calculation of the thermodynamic properties.\\

We also studied the geodesic deviation equation and the existence of circular geodesic trajectories to show that these orbits are stable in a range of radii depending on the parameter $\alpha$. Then, similarly to Schwarzschild and Reissner-Nördstrom black holes, there exist an Innermost Stable Circular Orbit (ISCO) with a radius that we calculated numerically from the geodesic deviation equation.\\

We also presented a new stationary, axisymmetric generalization of the Nashed-Capozziello solutions and calculated their horizon structure, Hawking temperature and ergosphere. The most interesting feature of these solutions is the angular dependence of the horizon radius and temperature, which imply an angle-dependent thermal radiation spectrum.

Finally, we want to conclude this work emphasizing that all the solutions studied here were obtained in the context of $f(R)$ gravity and due to the allowed range for the parameter $\alpha$, they cannot be reduced to general relativity. Hence, there are many astrophysical features that can be studied about these solutions which may be used to distinguish them from general relativity metrics.

\section*{Acknowledgements}
The authors acknowledge partial financial support from
Dirección de Investigación-Sede Bogotá, Universidad
Nacional de Colombia (DIB-UNAL) under Project
No. 50071 and Grupo de Astronomía, Astrofísica y
Cosmología-Observatorio Astronómico Nacional.

\end{document}